\documentclass[conference]{IEEEtran}

\usepackage{cite}
\usepackage{tikz}
\usepackage{amsmath}

\usepackage{pbox}
\usepackage{upquote}
\usepackage{url}
\usepackage{booktabs}
\usepackage{xspace}

\usepackage{breakurl}
\usepackage{multirow}
\usepackage{color, soul}
\usepackage{listings}
\usepackage[export]{adjustbox}
\usepackage{subfig}
\usepackage{tikz}
\usetikzlibrary{plotmarks}
\usepackage{filecontents}
\usepackage{xcolor}
\usepackage{enumitem}

\usepackage{amsmath,amssymb,amsfonts}
\usepackage{algorithmic}
\usepackage{graphicx}
\usepackage{textcomp}
\usepackage[font={small}]{caption}
\usepackage{wrapfig}
\usepackage{colortbl}
\usepackage{balance}
\usepackage{xtab}

\definecolor{dkgreen}{rgb}{0,0.6,0}
\definecolor{gray}{rgb}{0.5,0.5,0.5}
\definecolor{mauve}{rgb}{0.58,0,0.82}
\definecolor{editorGray}{rgb}{0.95, 0.95, 0.95}
\definecolor{editorOcher}{rgb}{1, 0.5, 0}
\definecolor{editorGreen}{rgb}{0,0.6,0}
\definecolor{darkgreen}{RGB}{0,90,90}
\definecolor{lightgray}{rgb}{0.95, 0.95, 0.95}
\definecolor{darkgray}{rgb}{0.4, 0.4, 0.4}
\definecolor{orange}{rgb}{1,0.45,0.13}		
\definecolor{olive}{rgb}{0.17,0.59,0.20}
\definecolor{brown}{rgb}{0.69,0.31,0.31}
\definecolor{purple}{rgb}{0.38,0.18,0.81}
\definecolor{lightblue}{rgb}{0.1,0.57,0.7}
\definecolor{lightred}{rgb}{1,0.4,0.5}
\definecolor{dkgreen}{rgb}{0,0.6,0}
\definecolor{gray}{rgb}{0.5,0.5,0.5}
\definecolor{mauve}{rgb}{0.58,0,0.82}
\definecolor{editorGray}{rgb}{0.95, 0.95, 0.95}
\definecolor{editorOcher}{rgb}{1, 0.5, 0}
\definecolor{editorGreen}{rgb}{0,0.6,0}

\definecolor{MyMaroon}{HTML}{D50000}
\definecolor{MyYellow}{HTML}{FFD600}
\definecolor{MyGreen}{HTML}{00C853}

\definecolor{MyTeal}{HTML}{009688}
\definecolor{MyOrange}{HTML}{FF5722}
\definecolor{MyGray}{HTML}{E0E0E0}
\definecolor{MyIndigo}{HTML}{3F51B5}
\definecolor{MyPurple}{HTML}{9C27B0}
\definecolor{MyBlack}{HTML}{000000}
\definecolor{MyBlue}{HTML}{3F51B5}

\definecolor{MyMaroon-f60000}{HTML}{f60000}
\definecolor{MyMaroon-ff3232}{HTML}{ff3232}
\definecolor{MyMaroon-ff8080}{HTML}{ff8080}
\definecolor{MyMaroon-ffbbbb}{HTML}{ffbbbb}
\definecolor{MyMaroon-fc6565}{HTML}{fc6565}
\lstdefinelanguage{JavaScript}{
  morekeywords={typeof, window, new, true, false, catch, function, return, null, catch, switch, var, if, in, else, case, break, document, write, createElement, width, height, display, visibility, border, document.write},
  morecomment=[s]{/*}{*/},
  morecomment=[l]//,
  morestring=[b]",
  morestring=[b]'
}

\lstdefinelanguage{HTML5}{
  language=html,
  sensitive=true,	
  alsoletter={<>=-},	
  morecomment=[s]{<!-}{-->},
  tag=[s],
  otherkeywords={
  >,
	<!DOCTYPE,
  </html, <html, <head, <title, </title, <style, </style, <link, </head, <meta, />,
	</body, <body,
	</div, <div, </div>,
	</p, <p, </p>, <a, <h1, </h1>, </h1,
	</script, <script,
  <canvas, /canvas>, <svg, <rect, <animateTransform, </rect>, </svg>, <video, <source, <iframe, </iframe>, </video>, <image, </image>, <header, </header, <article, </article, </iframe, <img, <span, </span, </a,<button, </button>, <ul>, </ul>, <li>, </li>
  },
  ndkeywords={
  =, ===, ==,
  charset=, src=, id=, width=, height=, style=, type=, rel=, href=, name=, tabindex=, align=, scrolling=, framespacing=, frameborder=, allowtransparency=, data-dm-title=, data-dm-format=, data-dm-filesize=, target=, data-dm=, data-dm-icon=, data-dm-href-free=, data-dm-filename=, data-dm-hosted-file=, data-dm-href= , data-dm-carregado=, class=, alt=,
  fill=, attributeName=, begin=, dur=, from=, to=, poster=, controls=, x=, y=, repeatCount=, xlink:href=,
  margin:, padding:, background-image:, border:, top:, left:, position:, width:, height:, margin-top:, margin-bottom:, font-size:, line-height:,
  transform:, -moz-transform:, -webkit-transform:,
  animation:, -webkit-animation:,
  transition:,  transition-duration:, transition-property:, transition-timing-function:,
  }
}

\lstdefinestyle{htmlcssjs} {%
  backgroundcolor=\color{editorGray},
  basicstyle=\fontsize{8}{8}\ttfamily,
  frame=tb,
  captionpos=b,
  belowcaptionskip=\medskipamount,
  xleftmargin={0.5cm},
  numbers=left,
  stepnumber=1,
  firstnumber=1,
  numberfirstline=true,	
  identifierstyle=\color{black},
  keywordstyle=\color{blue}\ttfamily,
  ndkeywordstyle=\color{editorGreen}\ttfamily,
  stringstyle=\color{black}\ttfamily,
  commentstyle=\color{brown}\ttfamily,
  language=HTML5,
  alsolanguage=JavaScript,
  alsodigit={.:;},	
  tabsize=2,
  showtabs=false,
  showspaces=false,
  showstringspaces=false,
  extendedchars=true,
  breaklines=true,
  numberstyle=\tiny\color{gray},
  literate=%
  {Ö}{{\"O}}1
  {Ä}{{\"A}}1
  {Ü}{{\"U}}1
  {ß}{{\ss}}1
  {ü}{{\"u}}1
  {ä}{{\"a}}1
  {ö}{{\"o}}1
}
\def\BibTeX{{\rm B\kern-.05em{\sc i\kern-.025em b}\kern-.08em
    T\kern-.1667em\lower.7ex\hbox{E}\kern-.125emX}}

\newcommand{\point}[1]{\vspace{.05in} \par\noindent\textbf{#1}.}

\definecolor{MyMaroon}{HTML}{D50000}
\definecolor{MyGreen}{HTML}{00C853}

\DeclareCaptionType{Payload}

\newcommand{\SuccesfullWebsitesCrawled}{17,629\xspace}
\newcommand{\DistinctExecutingScripts}{153,354\xspace}

\newcommand{\SuccesfullASTs}{143,526\xspace}
\newcommand{\FailedASTs}{9,828\xspace}
\newcommand{\FailedASTsPercent}{6.4\%\xspace}
\newcommand{\RawStaticFeatures}{47,717\xspace}
\newcommand{\VarinaceThreshold}{0.01\xspace}
\newcommand{\VarinaceThresholdFeatures}{8,597\xspace}
\newcommand{\RawDynamicFeatures}{2,628\xspace}

\newcommand{\TopTenKFingerprintingWebsites}{10.18\%\xspace}
\newcommand{\DormantDetectedByStatic}{94.46\%\xspace}
\newcommand{\ObfuscatedDetectedByDynamic}{92.30\%\xspace}

\newcommand{\PercentageMoreDetectionByClassifiers}{26\%\xspace}
\newcommand{\ClassifierFPR}{0.05\%\xspace}
\newcommand{\ClassifierFNR}{6.1\%\xspace}

\begin{document}

\newcommand{\framework}{\textsc{FP-Inspector}\xspace}

\title{\Large \bf Fingerprinting the Fingerprinters: \\ Learning to Detect Browser Fingerprinting Behaviors}

\author{\IEEEauthorblockN{Umar Iqbal}
\IEEEauthorblockA{\textit{The University of Iowa} }
\and
\IEEEauthorblockN{Steven Englehardt}
\IEEEauthorblockA{\textit{Mozilla Corporation} }
\and
\IEEEauthorblockN{Zubair Shafiq}
\IEEEauthorblockA{\textit{University of California, Davis} }
}

\maketitle

\begin{abstract}
Browser fingerprinting is an invasive and opaque stateless tracking technique. 
Browser vendors, academics, and standards bodies have long struggled to provide meaningful protections against browser fingerprinting that are both accurate and do not degrade user experience. 
We propose \framework, a machine learning based syntactic-semantic approach to accurately detect browser fingerprinting.
We show that \framework performs well, allowing us to detect 26\% more fingerprinting scripts than the state-of-the-art.
We show that an API-level fingerprinting countermeasure, built upon \framework, helps reduce website breakage by a factor of 2.
We use \framework to perform a measurement study of browser fingerprinting on top-100K websites.
We find that browser fingerprinting is now present on more than 10\% of the top-100K websites and over a quarter of the top-10K websites.
We also discover previously unreported uses of JavaScript APIs by fingerprinting scripts suggesting that they are looking to exploit  APIs in new and unexpected ways.
\end{abstract} 

\section{Introduction}
\label{sec:introduction}
Mainstream browsers have started to provide built-in protection against cross-site tracking.
For example, Safari \cite{safari_full_third_party_cookie_blocking} now blocks all third-party cookies and Firefox \cite{wood2019firefox69release} blocks third-party cookies from known trackers by default.
As mainstream browsers implement countermeasures against stateful tracking, there are concerns that it will encourage trackers to migrate to more opaque, stateless tracking techniques such as browser fingerprinting \cite{schuh2019building}.
Thus, mainstream browsers have started to explore mitigations for browser fingerprinting.

Some browsers and privacy tools have tried to mitigate browser fingerprinting by changing the JavaScript API surface exposed by browsers to the web.
For example, privacy-oriented browsers such as the Tor Browser \cite{laperdrix2019torfingerprinting,torBrowserFingerprintingBugs} have restricted access to APIs such as Canvas and WebRTC, that are known to be abused for browser fingerprinting.
However, such blanket API restriction has the side effect of breaking websites that use these APIs to implement benign functionality.

Mainstream browsers have so far avoided deployment of comprehensive API restrictions due to website breakage concerns.
As an alternative, some browsers---Firefox in particular \cite{eledstein2019firefoxfingerprinting}---have tried to mitigate browser fingerprinting by blocking network requests to browser fingerprinting services \cite{disconnect_me}.
However, this approach relies heavily on manual analysis and struggles to restrict fingerprinting scripts that are served from first-party domains or dual-purpose third parties, such as CDNs.
Englehardt and Narayanan \cite{Englehardt16MillionSiteMeasurementCCS} manually designed heuristics to detect fingerprinting scripts based on their execution behavior.
However, this approach relies on hard-coded heuristics that are narrowly defined to avoid false positives and must be continually updated to capture evolving fingerprinting and non-fingerprinting behaviors.

We propose \framework, a machine learning based approach to detect browser fingerprinting.
\framework trains classifiers to learn fingerprinting behaviors by extracting syntactic and semantic features through a combination of static and dynamic analysis that complement each others' limitations.
More specifically, static analysis helps \framework overcome the \emph{coverage} issues of dynamic analysis, while dynamic analysis overcomes the inability of static analysis to handle \emph{obfuscation}.

Our evaluation shows that \framework detects fingerprinting scripts with 99.9\% accuracy.
We find that \framework detects 26\% more fingerprinting scripts than manually designed heuristics \cite{Englehardt16MillionSiteMeasurementCCS}.
Our evaluation shows that \framework helps significantly reduce website breakage.
We find that targeted countermeasures that leverage \framework's detection reduce breakage by a factor 2 on websites that are particularly prone to breakage.

We deploy \framework to analyze the state of browser fingerprinting on the web.
We find that fingerprinting prevalence has increased over the years \cite{Acar14WebNeverForgetsCCS,Englehardt16MillionSiteMeasurementCCS}, and is now present on 10.18\% of the Alexa top-100K websites.
We detect fingerprinting scripts served from more than two thousand domains, which include both anti-ad fraud vendors as well as cross-site trackers. 
\framework also helps us uncover several new APIs that were previously not known to be used for fingerprinting.
We discover that fingerprinting scripts disproportionately use APIs such as the \texttt{Permissions} and \texttt{Performance} APIs.

We summarize our key contributions as follows:

\begin{enumerate}

\item An \textbf{ML-based syntactic-semantic approach} to detect browser fingerprinting behaviors by incorporating both static and dynamic analysis.

\item  An \textbf{evaluation of website breakage} caused by different mitigation strategies that block network requests or restrict APIs.

\item  A \textbf{measurement study} of browser fingerprinting scripts on the Alexa top-100K websites.

\item  A \textbf{clustering analysis of JavaScript APIs} to uncover new browser fingerprinting vectors.

\end{enumerate}

\noindent \textit{Paper Organization:} The rest of the paper proceeds as follows.
Section \ref{sec:background} presents an overview of browser fingerprinting and limitations of existing countermeasures.
Section \ref{sec:methodology} describes the design and implementation of \framework.
Section \ref{sec:evaluation} presents the evaluation of \framework's accuracy and website breakage.
Section \ref{sec:in-the-wild} describes \framework's deployment on Alexa top-100K websites.
Section \ref{sec:new-apis} presents the analysis of JavaScript APIs used by fingerprinting scripts.
Section \ref{sec:limitations} describes \framework's limitations.
Section \ref{sec:conclusion} concludes the paper.

\section{Background \& Related Work}
\label{sec:background}

\textbf{Browser fingerprinting for online tracking.}
Browser fingerprinting is a stateless tracking technique that uses device configuration information exposed by the browser through JavaScript APIs (e.g., \texttt{Canvas}) and HTTP headers (e.g., \texttt{User-Agent}).
In contrast to traditional stateful tracking, browser fingerprinting is stateless---the tracker does not need to store any client-side information (e.g., unique identifiers in cookies or local storage).
Browser fingerprinting is widely recognized by browser vendors \cite{mozillaFingerprintingBlogPost,privacyBudget,appleFingerprintingBlogPost} and standards bodies \cite{w3cFPGuidance,tagtracking} as an abusive practice.
Browser fingerprinting is more intrusive than cookie-based tracking for two reasons:
(1) while cookies are observable in the browser, browser fingerprints are opaque to users;
(2) while users can control cookies (e.g., disable third-party cookies or delete cookies altogether), they have no such control over browser fingerprinting.

Browser fingerprinting is widely known to be used for bot detection purposes \cite{BlueCava,Iovation,Threatmetrix,netiqDeviceFingerprintingWhitepaper}, including by Google's reCAPTCHA \cite{Bursztein16Picasso,Sivakorn16recaptcha} and during general web authentication \cite{Laperdrix19DIMVAMorellian,Alaca16ACSACDeviceFingerprinting}.
However, there are concerns that browser fingerprinting may be used for cross-site tracking especially as mainstream browsers such as Safari \cite{wilander2019itp23} and Firefox \cite{wood2019firefox69release} adopt aggressive policies against third-party cookies \cite{schuh2019building}.
For example, browser fingerprints (by themselves or when combined with IP address) \cite{laperdrix2019browser} can be used to regenerate or de-duplicate cookies \cite{TapadIdentityResolution,amperityIdentityResolution}.
In fact, as we show later, browser fingerprinting is used for both anti-fraud and potential cross-site tracking.

\textbf{Origins of browser fingerprinting.}
Mayer \cite{mayer2009pamphleteer} first showed that ``quirkiness'' can be exploited using JavaScript APIs (e.g., navigator, screen, Plugin, and MimeType objects) to identify users.
Later, Eckersley \cite{Eckersley10PETSuniquebrowser} conducted the Panopticlick experiment to analyze browser fingerprints using information from various HTTP headers and JavaScript APIs.
As modern web browsers have continued to add functionality through new JavaScript APIs \cite{snyder2016browser}, the browser's fingerprinting surface has continued to expand.
For example, researchers have shown that \texttt{Canvas} \cite{MoweryShacham2012PixelPerfect}, \texttt{WebGL} \cite{MoweryShacham2012PixelPerfect,Cao17CrossbrowserNDSS}, fonts \cite{Fifield2015FCfontmetrics}, extensions \cite{Starov2017XHound}, the \texttt{Audio} API \cite{Englehardt16MillionSiteMeasurementCCS}, the \texttt{Battery Status} API \cite{Olejnik15LeakingBattery}, and even mobile sensors \cite{Das18MobileSensorsCCS} can expose identifying device information that can be used to build a browser fingerprint.
In fact, many of these APIs have already been found to be abused in the wild \cite{nikiforakis2013cookieless,acar2013fpdetective,Acar14WebNeverForgetsCCS,Englehardt16MillionSiteMeasurementCCS,Das18MobileSensorsCCS,Olejnik17BatteryStatusIWPE}.
Due to these concerns, standards bodies such as the W3C \cite{W3CPrivacyWG} have provided guidance to take into account the fingerprinting potential of newly proposed JavaScript APIs.
One such example is the Battery Status API, which was deprecated by Firefox due to privacy concerns \cite{Olejnik17BatteryStatusIWPE}.

\textbf{Does browser fingerprinting provide unique and persistent identifiers?}
%
A browser fingerprint is a ``statistical'' identifier, meaning that it does not deterministically identify a device.
Instead, the identifiability of a device depends on the number of devices that share the same configuration.
Past research has reported widely varying statistics on the uniqueness of browser fingerprints.
Early research by Laperdrix et al. \cite{Laperdrix16BeautyAndTheBeastSP} and Eckersley \cite{Eckersley10PETSuniquebrowser} found that 83\% to 90\% of devices have a unique fingerprint.
In particular, Laperdrix et al. found that desktop browser fingerprints are more unique (90\% of devices) than mobile (81\% of devices) due to the presence of plugins and extensions.
However, both Eckersley's and Laperdrix's studies are based on data collected from self-selected audiences---visitors to Panopticlick and AmIUnique, respectively---which may bias their findings.
In a more recent study, Boix et al. \cite{Boix18HidingInTheCrowdWWW} deployed browser fingerprinting code on a major French publisher's website.
They found that only 33.6\% of the devices in that sample have unique fingerprints.
However, they argued that adding other properties, such as the IP address, \textit{Content language} or \textit{Timezone}, may make the fingerprint unique.

To be used as a tracking identifier, a browser fingerprint must either remain stable over time or be linkable with relatively high confidence.
Eckersley measured repeat visits to the Panopticlick test page and found that 37\% of repeat visitors had more than one fingerprint \cite{Eckersley10PETSuniquebrowser}.
However, about 65\% of devices could be re-identified by linking fingerprints using a simple heuristic.
Similarly, Vastel et al. \cite{vastel2018fp} found that half of the repeat visits to the AmIUnique test page change their fingerprints in less than 5 days.
They improve on Eckersley's linking heuristic and show that their linking technique can track repeat AmIUnique visitors for an average of 74 days.

\textbf{Prevalence of browser fingerprinting.}
A 2013 study of browser fingerprinting in the wild \cite{nikiforakis2013cookieless} examined three fingerprinting companies and found only 40 of the Alexa top-10K websites deploying fingerprinting techniques.
That same year, a large-scale study by Acar et al. \cite{acar2013fpdetective} found just 404 of the Alexa top 1-million websites deploying fingerprinting techniques.
Following that, a number of studies have measured the deployment of fingerprinting across the web \cite{Acar14WebNeverForgetsCCS,Englehardt16MillionSiteMeasurementCCS,Das18MobileSensorsCCS,Olejnik17BatteryStatusIWPE}.
Although these studies use different methods to fingerprinting, their results suggest an overall trend of increased fingerprinting deployment.
Most recently, an October 2019 study by The Washington Post \cite{fowlerWashingtonPostFingerprinting} found fingerprinting on about 37\% of the Alexa top-500 US websites.
This roughly aligns with our findings in Section~\ref{sec:in-the-wild}, where we discover fingerprinting scripts on 30.60\% of the Alexa top-1K websites.
Despite increased scrutiny by browser vendors and the public in general, fingerprinting continues to be prevalent.

\textbf{Browser fingerprinting countermeasures.}
Existing tools for fingerprinting protection broadly use three different approaches.\footnote{Google has recently proposed a new approach to fingerprinting protection that doesn't fall into the categories discussed above. They propose assigning a ``privacy cost'' based on the entropy exposed by each API access and enforcing a ``privacy budget'' across all API accesses from a given origin \cite{privacyBudget}. Since this proposal is only at the ideation stage and does not have any implementations, we do not discuss it further.}
One approach randomizes return values of the JavaScript APIs that can be fingerprinted, the second normalizes the return values of the JavaScript APIs that can be fingerprinted, and the third uses heuristics to detect and block fingerprinting scripts.
All of these approaches have different strengths and weaknesses.
Some approaches protect against \textit{active} fingerprinting, i.e. scripts that probe for device properties such as the installed fonts, and others protect against \textit{passive} fingerprinting, i.e. servers that collect information that's readily included in web requests, such as the \texttt{User-Agent} request header.
Randomization and normalization approaches can defend against all forms of active fingerprinting and some forms of passive (e.g., by randomizing \texttt{User-Agent} request header).
Heuristic-based approaches can defend against both active and passive fingerprinting, e.g., by completely blocking the network request to resource that fingerprints.
We further discuss these approaches and list their limitations.

\begin{enumerate}

\item The \emph{randomization} approaches, such as Canvas Defender \cite{canvas_defender}, randomize the return values of the APIs such as \texttt{Canvas} by adding noise to them.
These approaches not only impact the functional use case of APIs but are also ineffective at restricting fingerprinting as they are reversible \cite{Vestel18FpScannerUSENIX}.
Additionally, the noised outputs themselves can sometimes serve as a fingerprint, allowing websites to identify the set of users that have the protection enabled \cite{datta2018effectiveness, Vestel18FpScannerUSENIX}.

\item The \emph{JavaScript API normalization} approaches, such as those used by the Tor Browser \cite{TOR_FIngerprintProtection} and the Brave browser \cite{Brave_FingerprintingProtectionMode}, attempt to make all users return the same fingerprint.
This is achieved by limiting or spoofing the return values of some APIs (e.g., \texttt{Canvas}), and entirely removing access to other APIs (e.g., \texttt{Battery Status}).
These approaches limit website functionality and can cause websites to break, even when those websites are using the APIs for benign purposes.

\item The \emph{heuristic} approaches, such as Privacy Badger \cite{privacybadger_web} and Disconnect \cite{disconnect_me}, detect fingerprinting scripts with pre-defined heuristics.
Such heuristics, which must narrowly target fingerprinters to avoid over-blocking, have two limitations.
First, they may miss fingerprinting scripts that do not match their narrowly defined detection criteria.
Second, the detection criteria must be constantly maintained to detect new or evolving fingerprinting scripts.

\end{enumerate}

\textbf{Learning based solutions to detect fingerprinting.}
The ineffectiveness of randomization, normalization, and heuristic-based approaches motivate the need of a learning-based solution.
Browser fingerprinting falls into the broader class of \textit{stateless} tracking, i.e., tracking without storing on data on the user's machine.
Stateless tracking is in contrast to \textit{stateful} tracking, which uses APIs provided by the browser to store an identifier on the user's device.
Prior research has extensively explored learning-based solutions for detecting stateful trackers.
Such approaches try to learn tracking behavior of scripts based on their structure and execution.
One such method by Ikram et al. \cite{Ikram17SeamlessTrackingPETS} used features extracted through static code analysis.
They extracted n-grams of code statements as features and trained a one-class machine learning classifier to detect tracking scripts.
In another work, Wu et al. \cite{Wu16MLTrackingESORICS} used features extracted through dynamic analysis.
They extracted one-grams of web API method calls from execution traces of scripts as features and trained a machine learning classifier to detect tracking scripts.

Unfortunately, prior learning-based solutions generally lump together stateless and stateful tracking. 
However, both of these tracking techniques fundamentally differ from each other and a solution that tries to detect both stateful and stateless techniques will have mixed success.
For example, a recent graph-based machine learning approach to detect ads \emph{and} trackers proposed by Iqbal et al. \cite{Iqbal20AdGraphSP} at times successfully identified fingerprinting and at times failed.

Fingerprinting detection has not received as much attention as stateful tracking detection.
Al-Fannah et. al. \cite{AlFannah18CookieMonsterISC} proposed to detect fingerprinting vendors by matching 17 manually identified attributes (e.g., \texttt{User-Agent}), that have fingerprinting potential, with the request URL.
The request is labeled as fingerprinting if at least one of the attributes is present in the URL.
However, this simple approach would incorrectly detect the functional use of such attributes as fingerprinting. 
Moreover, this approach fails when the attribute values in the URL are hashed or encrypted.
Rizzo \cite{Valentino18MLFingerprintingThesis}, in their thesis, explored the detection of fingerprinting scripts using machine learning.
Specifically, they trained a machine learning classifier with features extracted through static code analysis.
However, only relying on static code analysis might not be sufficient for an effective solution.
Static code analysis has inherent limitations to interpret obfuscated code and provide clarity in enumerations.
These limitations may hinder the ability of a classifier, trained on features extracted through static analysis, to correctly detect fingerprinting scripts as both obfuscation \cite{Skolka19minifiedobfuscatedJS} and enumerations (canvas font fingerprinting) are common in fingerprinting scripts.
Dynamic analysis of fingerprinting scripts could solve that problem but it requires scripts to execute and scripts may require user input or browser events to trigger.

A complementary approach that uses both static and dynamic analysis could work---indeed this is the approach we take next in Section~\ref{sec:methodology}.
Dynamic analysis can provide interpretability for obfuscated scripts and scripts that involve enumerations and static analysis could provide interpretability for scripts that require user input or browser triggers.

\section{\framework}
\label{sec:methodology}
\begin{figure*}[!htp]
    \centering
    \includegraphics[width=\textwidth,trim=0.5cm 1.0cm 0.5cm 1.2cm]{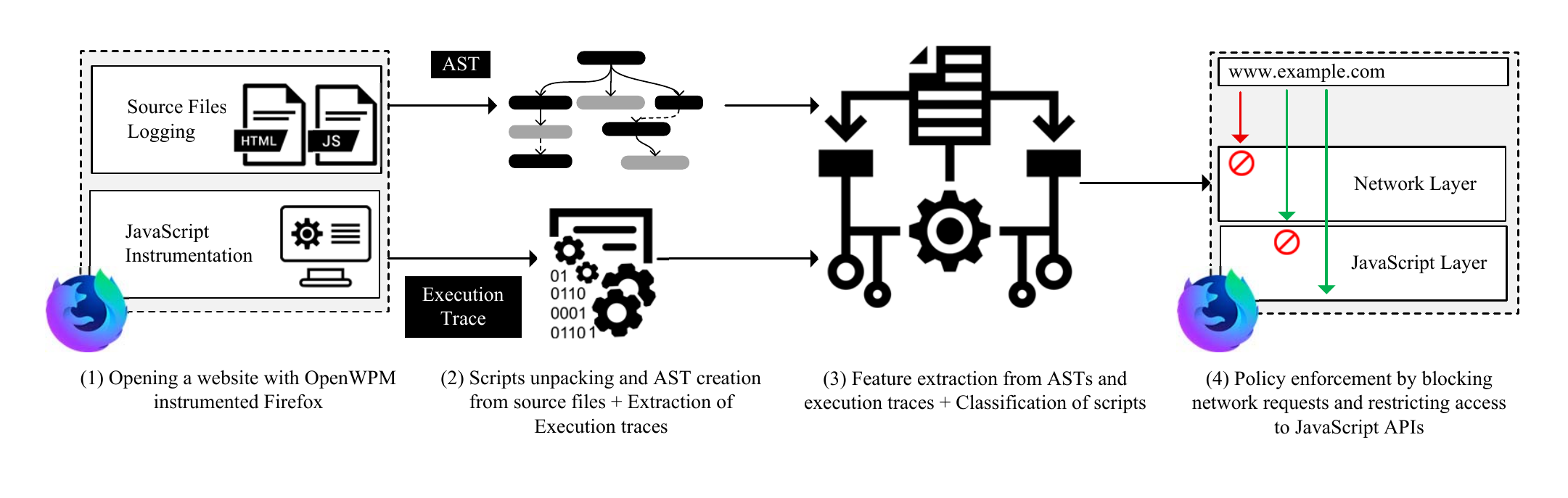}
    \caption{\framework:
    (1) We crawl the web with an extended version of OpenWPM that extracts JavaScript source files and their execution traces. (2) We extract Abstract Syntax Trees (ASTs) and execution traces for all scripts. (3) We use those representations to extract features and train a machine learning model to detect fingerprinting scripts. (4) We use a layered approach to counter fingerprinting scripts.}
\label{figure:approach_overview}
 \vspace{-10pt}
\end{figure*}

In this section we present the design and implementation of \framework, a machine learning approach that combines static and dynamic JavaScript analysis to counter browser fingerprinting.
\framework has two major components:
the \textit{detection component}, which extracts syntactic and semantic features from scripts and trains a machine learning classifier to detect fingerprinting scripts; and
the \textit{mitigation component}, which applies a layered set of restrictions to the detected fingerprinting scripts to counter passive and/or active fingerprinting in the browser.
Figure~\ref{figure:approach_overview} summarizes the architecture of \framework.

\subsection{Detecting fingerprinting scripts}
\label{subsec: detecting}
A fingerprinting script has a limited number of APIs it can use to extract a specific piece of information from a device.
For example, a script that tries to inspect the graphics stack must use the \texttt{Canvas} and \texttt{WebGL} APIs; if a script wants to collect 2D renderings (i.e., for canvas fingerprinting), it must call \texttt{toDataURL()} or \texttt{getImageData()} functions of the Canvas API to access the rendered canvas images.
Past research has used these patterns to manually curate heuristics for detecting fingerprinting scripts with fairly high precision \cite{Englehardt16MillionSiteMeasurementCCS,Das18MobileSensorsCCS}. 
Our work builds on them and significantly extends prior work in two main ways.

First, \framework \emph{automatically} learns emergent properties of fingerprinting scripts instead of relying on hand-coded heuristics.
Specifically, we extract a large number of low-level heuristics for capturing syntactic and semantic properties of fingerprinting scripts to train a machine learning classifier.
\framework's classifier trained on limited ground truth of fingerprinting scripts from prior research is able to generalize to detect new fingerprinting scripts as well as previously unknown fingerprinting methods.

Second, unlike prior work, we leverage \emph{both} static features (i.e., script syntax) and dynamic features (i.e., script execution).
The static representation allows us to capture fingerprinting scripts or routines that may not execute during our page visit (e.g., because they require user interaction that is hard to simulate during automated crawls).
The dynamic representation allows us to capture fingerprinting scripts that are obfuscated or minified.
\framework trains separate supervised machine learning models for static and dynamic representations and combines their output to accurately classify a script as fingerprinting or non-fingerprinting.

\textbf{Script monitoring.}
We gather script contents and their execution traces by automatically loading webpages in an extended version of OpenWPM \cite{Englehardt16MillionSiteMeasurementCCS}. 
By collecting both the raw content and dynamic execution traces of scripts, we are able to use both static and dynamic analysis to extract features related to fingerprinting.

\textit{Collecting script contents:}
We collect script contents by extending OpenWPM's network monitoring instrumentation.
By default, this instrumentation saves the contents of all HTTP responses that are loaded into script tags.
We extend OpenWPM to also capture the response content for all HTML documents loaded by the browser.
This allows us to capture both external and inline JavaScript.
We further parse the HTML documents to extract inline scripts.
This detail is crucial because a vast majority of webpages use inline scripts \cite{Nikiforakis12jsinclusion,Lauinger17JSlibraries}.

\textit{Collecting script execution traces:}
We collect script execution traces by extending OpenWPM's script execution instrumentation.
OpenWPM records the name of the Javascript API being accessed by a script, the method name or property name of the access, any arguments passed to the method or values set or returned by the property, and the stack trace at the time of the call.
By default, OpenWPM only instruments a limited number of the JavaScript APIs that are known to be used by fingerprinting scripts.
We extend OpenWPM script execution instrumentation to cover additional APIs and script interactions that we expect to provide useful information for differentiating fingerprinting activity from non-fingerprinting activity.
There is no canonical list of fingerprintable APIs, and it is not performant to instrument the browser's entire API surface within OpenWPM.
In light of these constraints, we extended the set of APIs instrumented by OpenWPM to cover several additional APIs used by popular fingerprinting libraries (i.e., fingerprintjs2 \cite{Fingerprint2js}) and scripts (i.e., MediaMath's fingerprinting script \cite{mathtag}).\footnote{The full set of APIs monitored by our extended version of OpenWPM in Appendix~\ref{app:js-extensions}.}
These include the Web Graphics Library (\texttt{WebGL}) and \texttt{performance.now}, both of which were previously not monitored by OpenWPM.
We also instrument a number of APIs used for Document Object Model (DOM) interactions, including the \texttt{createElement} method and the \texttt{document} and \texttt{node} objects.
Monitoring access to these APIs allows us to differentiate between scripts that interact with the DOM and those that do not.

\textbf{Static analysis.}
Static analysis allows us to capture information from the contents and structure of JavaScript files---including those which did not execute during our measurements or those which were not covered by our extended instrumentation.

\textit{AST representation:}
First, we represent scripts as Abstract Syntax Trees (ASTs).
This allows us to ignore coding style differences between scripts and ever changing JavaScript syntax.
ASTs encode scripts as a tree of syntax primitives (e.g., \texttt{VariableDeclaration} and \texttt{ForStatement}), where edges represent syntactic relationship between code statements.
If we were to build features directly from the raw contents of scripts, we would encode extraneous information that may make it more difficult to determine whether a script is fingerprinting.
As an example, one script author may choose to loop through an array of device properties by index, while another may choose to use that same array's \texttt{forEach} method.
Both scripts are accessing the same device information in a loop, and both scripts will have a similar representation when encoded as ASTs.

\begin{figure}[h]
      \centering
          \subfloat[AST for packed Script \ref{lst:sample_fp_script_packed}]{
          \includegraphics[width=0.35\textwidth]{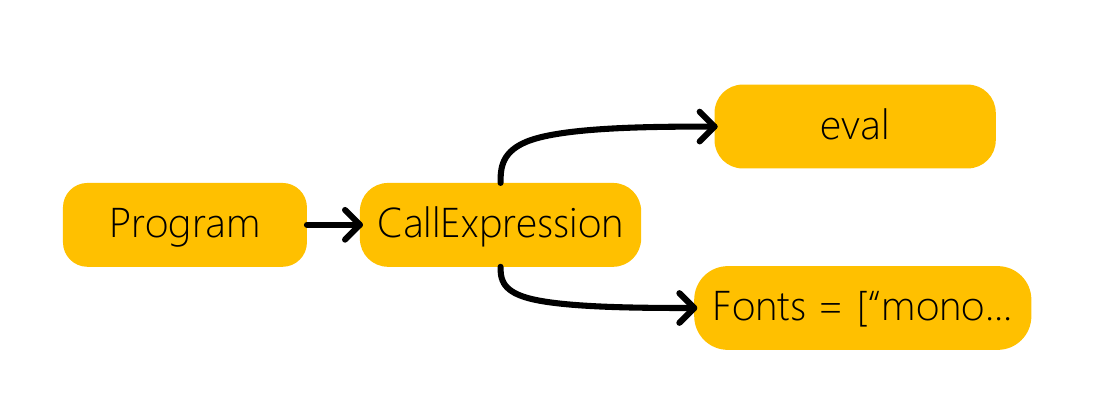}
          \label{figure:example_sub_ast_packed}
          }
           \vspace{-17pt}
          \hfill
           \subfloat[AST for unpacked script \ref{lst:sample_fp_script}]{
          \includegraphics[width=0.4\textwidth]{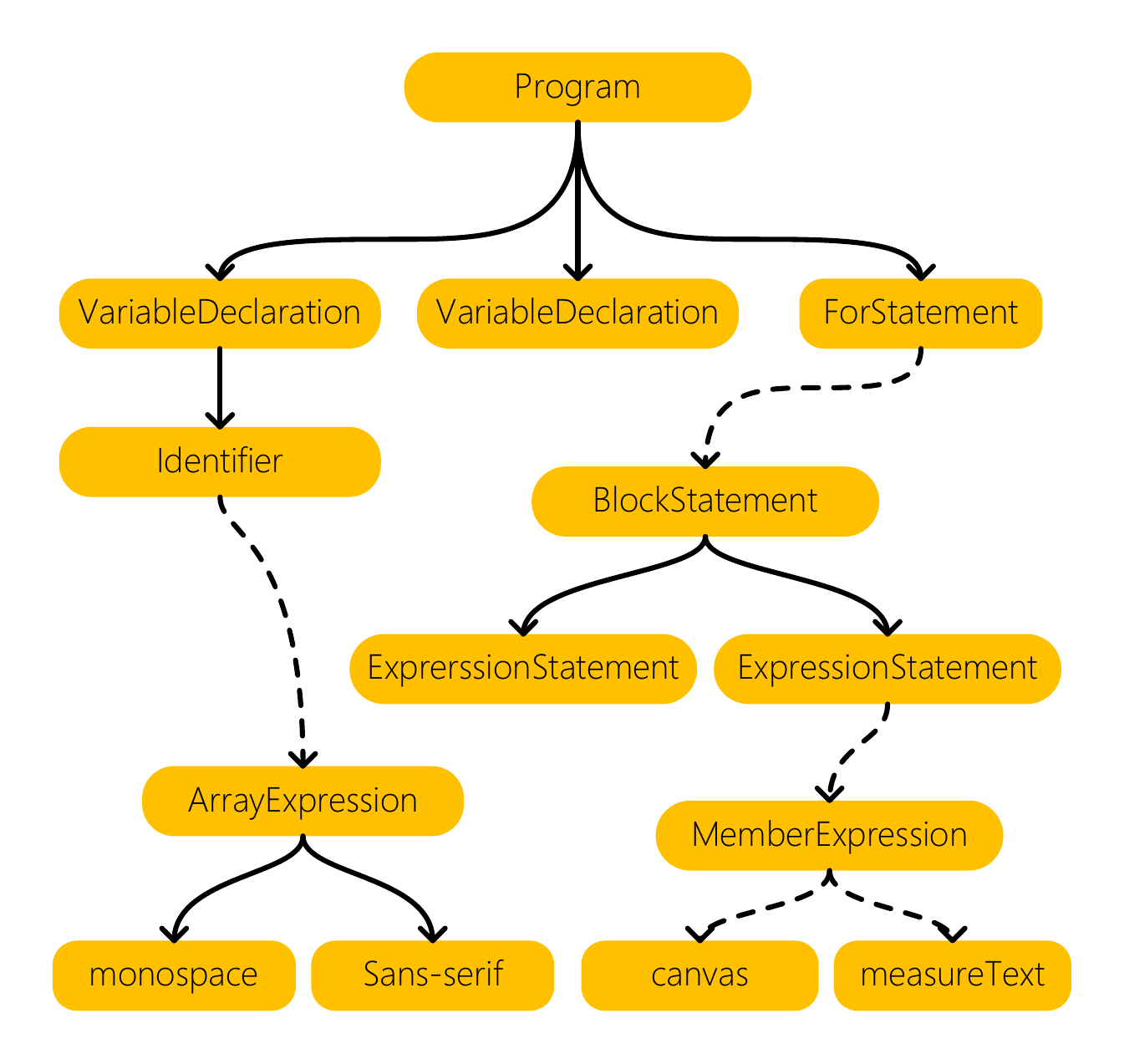}
          \label{figure:example_sub_ast}
          }
      \caption{A truncated AST representation of Scripts \ref{lst:sample_fp_script_packed} and \ref{lst:sample_fp_script}. The edges represent the syntactic relationship between nodes. Dotted lines indicate an indirect connection through truncated nodes.}
      \label{figure:example_ast}
       \vspace{-12pt}
\end{figure}

Figure \ref{figure:example_sub_ast} provides an example AST built from a simple script.
Nodes in an AST represent keywords, identifiers, and literals in the script, while edges represent the relation between them.
\textit{Keywords} are reserved words that have a special meaning for the interpreter (e.g. \texttt{for}, \texttt{eval}), \textit{identifiers} are function names or variable names (e.g. \texttt{CanvasElem}, \texttt{FPDict}), and \textit{literals} are constant values, such as a string assigned to an identifier (e.g. ``\texttt{example}'').
Note that whitespace, comments, and coding style are abstracted away by the AST.

\textit{Script unpacking:}
The process of representing scripts as ASTs is complicated by the fact that JavaScript is an interpreted language and compiled at run time.
This allows portions of the script to arrive as plain text which is later compiled and executed with \texttt{eval} or \texttt{Function}.
Prior work has shown that the fingerprinting scripts often include code that has been ``packed'' with \texttt{eval} or \texttt{Function} \cite{Skolka19minifiedobfuscatedJS}.
To unpack scripts containing \texttt{eval} or \texttt{Function}, we embed them in empty HTML webpages and open them in an instrumented browser \cite{Iqbal20AdGraphSP} which allows us to extract scripts as they are parsed by the JavaScript engine.
We capture the parsed scripts and use them in place of the packed versions when building ASTs.
We also follow this same procedure to extract in-line scripts, which are scripts included directly in the HTML document.

Script~\ref{lst:sample_fp_script_packed} shows an example canvas font fingerprinting script that has been packed with \texttt{eval}.
This script loops through a list of known fonts and measures the rendered width to determine whether the font is installed (see \cite{Englehardt16MillionSiteMeasurementCCS} for a thorough description of canvas font fingerprinting).
Script~\ref{lst:sample_fp_script} shows the unpacked version of the script.
As can be seen from the two snippets, the script is significantly more interpretable after unpacking.
Figure~\ref{figure:example_ast} shows the importance of unpacking to AST generation.
The packed version of the script (i.e., Script~\ref{lst:sample_fp_script_packed}) creates a generic stub AST (i.e., Figure~\ref{figure:example_sub_ast_packed}) which would match the AST of any script that uses \texttt{eval}.
Figure~\ref{figure:example_sub_ast} shows the full AST that has been generated from the unpacked version of the script (i.e., Script~\ref{lst:sample_fp_script}).
This AST captures the actual structure and content of the fingerprinting code that was passed to \texttt{eval}, and will allow us to extract meaningful features from the script's contents.

\begin{figure}[!tpb]
\begin{lstlisting}[basicstyle=\linespread{0.2},style=htmlcssjs,caption=A canvas font fingerprinting script packed with eval., label={lst:sample_fp_script_packed}]
eval("Fonts =[\"monospace\",..,\"sans-serif\"];
CanvasElem = document.createElement(\"canvas\")
;CanvasElem.width = \"100\";CanvasElem.height =
\"100\";context = CanvasElem.getContext('2d');
FPDict= {};for(i=0;i<Fonts.length;i++){
CanvasElem.font = Fonts[i];FPDict[Fonts[i]] =
CanvasElem.measureText(\"example\").width;}")
\end{lstlisting}
 \vspace{-20pt}
\end{figure}

\begin{figure}[!tpb]
\begin{lstlisting}[basicstyle=\linespread{0.2},style=htmlcssjs,caption=An unpacked version of the script in Script~\ref{lst:sample_fp_script_packed}., label={lst:sample_fp_script}]
// Canvas font fingerprinting script.
Fonts = ["monospace" , ... , "sans-serif"];

CanvasElem = document.createElement("canvas");
CanvasElem.width = "100";
CanvasElem.height = "100";
context = CanvasElem.getContext('2d');
FPDict= {};
for (i = 0; i < Fonts.length; i++)
{
  CanvasElem.font = Fonts[i];
  FPDict[Fonts[i]] = context.measureText("example").width;
}
\end{lstlisting}
 \vspace{-25pt}
\end{figure}

\textit{Static feature extraction:}
Next, we generate static features from ASTs.
ASTs have been extensively used in prior research to detect malicious JavaScript \cite{Curtsinger11ZozzleUSENIX,Iqbal17AntiABIMC,Fass2019ACSACJStrap}.
To build our features, we first hierarchically traverse the ASTs and divide them into pairs of parent and child nodes.
Parents represents the context (e.g., \texttt{for} loops, \texttt{try} statements, or \texttt{if} conditions), and children represent the function inside that context (e.g., \texttt{createElement}, \texttt{toDataURL}, and \texttt{measureText}).
Naively parsing \texttt{parent:child} pairs for the entire AST of every script would result in a prohibitively large number of features across all scripts (i.e., millions).
To avoid this we only consider \texttt{parent:child} pairs that contain at least one keyword that matches a name, method, or property from one of the JavaScript APIs \cite{MDNWebAPIs}.
We assemble these \texttt{parent:child} combinations as feature vectors for all scripts.
Each \texttt{parent:child} combination is treated as a binary feature, where 1 indicates the presence of a feature and 0 indicates its absence.
Since we do not execute scripts in static analysis, fingerprinting-specific JavaScript API methods usually have only one occurrence in the script.
Thus, we found the binary representation to sufficiently capture this information from the script.

As an example, feature extracted from AST in Figure \ref{figure:example_sub_ast} have \texttt{ForStatement:var} and \texttt{MemberExpression:measureText} as features which indicate the presence of a loop and access to \texttt{measureText} method.
These methods are frequently used in canvas font fingerprinting scripts.
Intuitively, fingerprinting script vectors have combinations of \texttt{parent:child} pairs that are specific to an API access pattern indicative of fingerprinting (e.g., setting a new font and measuring its width within a loop) that are unlikely to occur in non-fingerprinting scripts.
A more comprehensive list of features extracted from the AST in Figure \ref{figure:example_sub_ast} are listed in Appendix \ref{app:sample-features} (Table \ref{table:static_features_list}).

To avoid over-fitting, we apply unsupervised and supervised feature selection methods to reduce the number of features.
Specifically, we first prune features that do not vary much (i.e., variance $<$ 0.01) and also use information gain \cite{QuinlanInformationGain} to short list top-1K features.
This allows us to keep the features that represent the most commonly used APIs for fingerprinting.
For example, two of the features with the highest information gain represent the usage of \texttt{getSupportedExtensions} and \texttt{toDataURL} APIs.
\texttt{getSupportedExtensions} is used to get the list of supported WebGL extensions, which vary depending on browser's implementation.
\texttt{toDataURL} is used to get the base64 representation of the drawn canvas image, which depending on underlying hardware and OS configurations differs for the same canvas image.
We then use these top-1K features as input to train a supervised machine learning model.

\begin{figure}[!ht]
	\centering
    \subfloat[Obfuscated canvas font fingerprinting script from Script~\ref{lst:sample_fp_script}.\label{figure:sample_fp_script_obfuscated}]{
    	\parbox{.48\textwidth}{
\begin{lstlisting}[style=htmlcssjs]
var _0x2c4a=['\x63\x58\x49\x69','\x42\x6a\x58\
x44\x6f\x41\x3d\x3d','\x55\x54\x72\x43\x69\x73
\x4f\x77\x4f\x38\x4f\x6c\x50\x45\x6e\x43\x6d\x
77\x30\x3d','\x49\x38\x4f\x38\x49\x4d\x4f\x42\
x77\x70\x72\x44\x6e\x41\x3d\x3d','\x77\x35\x54
\x43\x73\x42\x56\x51','\x77\x37\x62\x43\x69\x4
d\x4f\x38\x77\x...............................
....................................x3284af={};
for(i=0x0;i<_0x1b2b65[_0x5d52('0x7','\x28\x6d\
x68\x26')];i++){_0x1d1d56[_0x5d52('0x8','\x67\
x33\x48\x21')]=_0x1b2b65[i];_0x3284af[_0x1b2b6
5[i]]=_0x4d24cc[_0x5d52('0x9','\x35\x70\x64\x4
c')](_0x5d52('0xa','\x28\x6d\x68\x26'))['\x77\
x69\x64\x74\x68'];}
\end{lstlisting}
    	}
    }
               \vspace{-10pt}
    \hfill
    \subfloat[AST of the obfuscated script shown in (a).\label{figure:example_sub_ast_obfuscated}]{
        \parbox{.45\textwidth}{
        	\includegraphics[width=0.4\textwidth,valign=t]{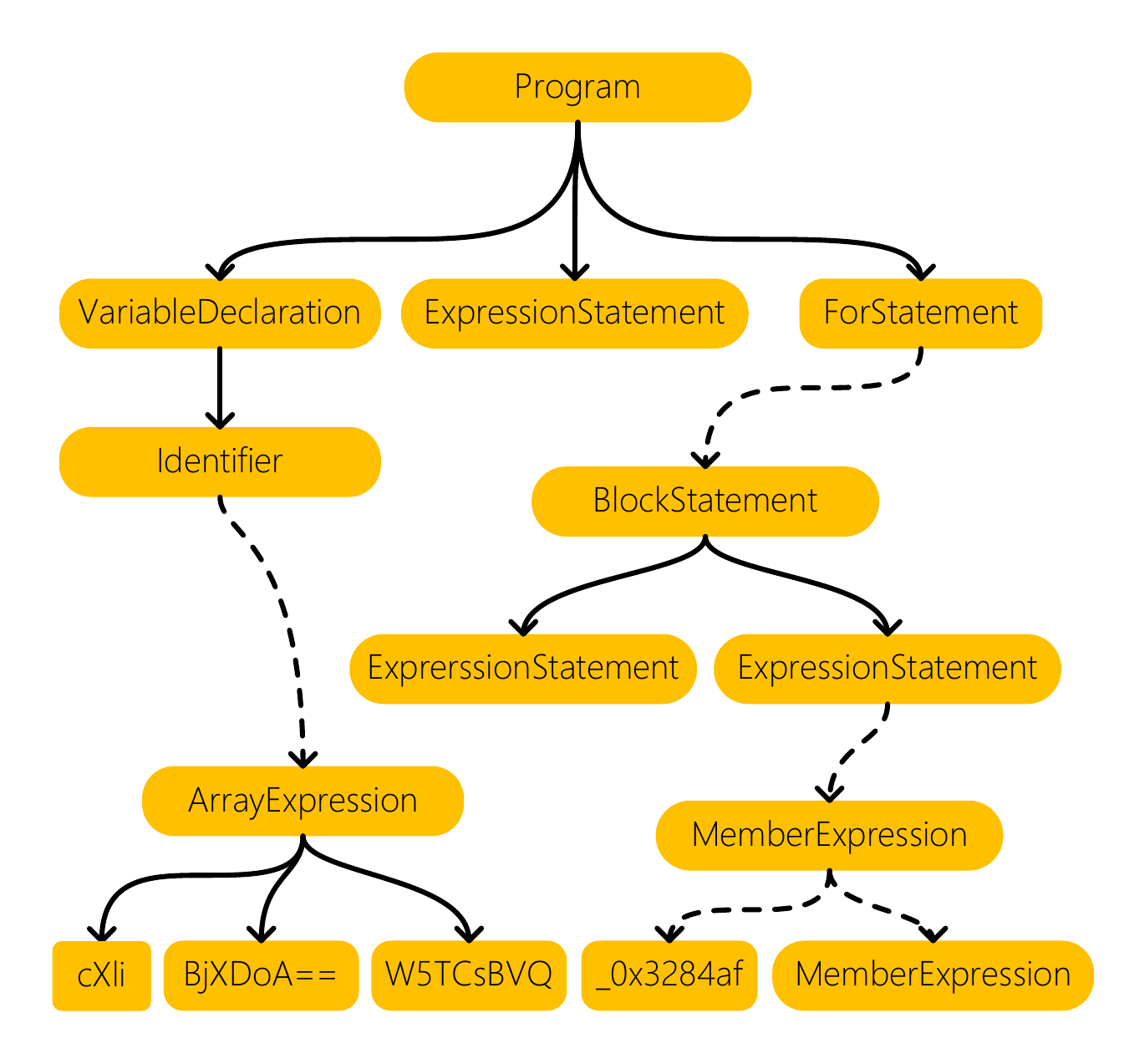}
        }
    }
    \caption{A truncated example showing the AST representation of an obfuscated version of the canvas font fingerprinting script in Script~\ref{lst:sample_fp_script}. The edges represent the syntactic relationship between nodes. Dotted lines indicate an indirect connection through truncated nodes.}
    \label{fig:obfuscated-ast}
     \vspace{-15pt}
\end{figure}

\textbf{Dynamic analysis.}
Dynamic analysis complements some weaknesses of static analysis.
While static analysis allows us to capture the syntactic structure of scripts, it fails when the scripts are obfuscated or minified.
This is crucial because prior research has shown that fingerprinting scripts often use obfuscation to hide their functionality \cite{Skolka19minifiedobfuscatedJS}.
For example, Figure~\ref{fig:obfuscated-ast} shows an AST constructed from an obfuscated version of Script \ref{lst:sample_fp_script}.
The static features extracted from this AST would miss important \texttt{parent:child} pairs that are essential to capturing the script's functionality.
Furthermore, some of the important \texttt{parent:child} pairs may be filtered during feature selection.
Thus, in addition to extracting static features from script contents, we extract dynamic features by monitoring the execution of scripts.
Execution traces capture the semantic relationship within scripts and thus provide additional context regarding a script's functionality, even when that script is obfuscated.

\textit{Dynamic feature extraction:}
We use two approaches to extract features from execution traces.
First, we keep presence and count of the number of times a script accesses each individual API method or property and use that as a feature.
Next, we build features from APIs that are passed arguments or return values.
Rather than using the arguments or return values directly, we use derived values to capture a higher-level semantic that is likely to better generalize during classification.
For example, we will compute the length of a string rather than including the exact text, or will compute the area of a element rather than including the height and width.
This allows us to avoid training our classifier with overly specific features---i.e., we do not care whether the text ``CanvasFingerprint'' or ``C4NV45F1NG3RPR1NT'' is used during a canvas fingerprinting attempt, and instead only care about the text length and complexity.
For concrete example, we calculate the area of canvas element, its text size, and whether its is present on screen when processing execution logs related to \texttt{CanvasRenderingContext2D.fillText()}.

As an example, the features extracted from the execution trace of Script \ref{figure:sample_fp_script_obfuscated} includes (\texttt{HTMLCanvasElement.getContext, True}) and (\texttt{CanvasRenderingContext2D.measureText, 7}) as features, where \texttt{True} indicates the usage of \texttt{HTMLCanvasElement.getContext} and \texttt{7} indicates the size of text in \texttt{CanvasRenderingContext2D.measureText}.
A more comprehensive list of features extracted from the execution trace of Script \ref{figure:sample_fp_script_obfuscated} can be found in Appendix \ref{app:sample-features} (Table \ref{table:dynamic_features_list}).

To avoid over-fitting, we again apply unsupervised and supervised feature selection methods to limit the number of features.
%
%
Similar to feature reduction for static analysis, this allows us to keep the features that represent the most commonly used APIs for fingerprinting.
For example, two of the features with the highest information gain represent the usage of \texttt{CanvasRenderingContext2D.fillStyle} and \texttt{navigator.platform} APIs.
\texttt{CanvasRenderingContext2D.fillStyle} is used to specify the color, gradient, or pattern inside a canvas shape, which can make a shape render differently across browsers and devices.
\texttt{navigator.platform} reveals the platform (e.g. MacIntel and Win32) on which the browser is running.
We then use these top-1K features as input to train a supervised machine learning model.

\textbf{Classifying fingerprinting scripts.}
\framework uses a decision tree \cite{Quinlan1993DecisionTree} classifier for training a machine learning model.
The decision tree is passed feature vectors of scripts for classification.
While constructing the tree, at each node, the decision tree chooses the feature that most effectively splits the data.
Specifically, the attribute with highest information gain is chosen to split the data by enriching one class.
The decision tree then follows the same methodology to recursively partition the subsets unless the subset belongs to one class or it can no longer be partitioned.

Note that we train two separate models and take the union of their classification results instead of combining features from both the static and dynamic representations of scripts to train a single model.
That is, a script is considered to be a fingerprinting script if it is classified as fingerprinting by either the model that uses static features as input or the model that uses dynamic features as input.
We use union of the two models because we only have the decision from one of the two models for some scripts (e.g., scripts that do not execute).
Furthermore, the two models are already trained on high-precision ground truth \cite{Englehardt16MillionSiteMeasurementCCS} and taking the union would allow us to push for better recall.
Using this approach, we classify all scripts loaded during a page visit---i.e., we include both external scripts loaded from separate URLs and inline scripts contained in any HTML document.

\subsection{Mitigating fingerprinting scripts}
\label{subsec: mitigating}
Existing browser fingerprinting countermeasures can be classified into two categories: content blocking and API restriction.
Content blocking, as the name implies, blocks the requests to download fingerprinting scripts based on their network location (e.g., domain or URL).
API restriction, on the other hand, does not block fingerprinting scripts from loading but rather limits access to certain JavaScript APIs that are known to be used for browser fingerprinting.

Privacy-focused browsers such as the Tor Browser \cite{TOR_FIngerprintProtection} prefer blanket API restriction over content blocking mainly because it side steps the challenging problem of detecting fingerprinting scripts.
While API restriction provides reliable protection against active fingerprinting, it can break the functionality of websites that use the restricted APIs for benign purposes.
Browsers that deploy API restriction also require additional protections against passive fingerprinting (e.g., routing traffic over the Tor network).
Content blocking protects against both active and passive fingerprinting, but it is also prone to breakage when the detected script is dual-purpose (i.e., implements both fingerprinting and legitimate functionality) or a false positive.

Website breakage is an important consideration for fingerprinting countermeasures.
For instance, a recent user trial by Mozilla showed that privacy countermeasures in Firefox can negatively impact user engagement due to website breakage \cite{MozillaBreakageUserStudy}.
In fact, website breakage can be the deciding factor in real-world deployment of any privacy-enhancing countermeasure \cite{MozillaPostpone3PCookieBlocking,MozilaDefaultCookieRestriction}.
We are interested in studying the impact of different fingerprinting countermeasures based on \framework on website breakage.
We implement the following countermeasures:

\begin{enumerate}[wide, labelwidth=!, labelindent=0pt]

\item \textbf{Blanket API Restriction.} We restrict access for all scripts to the JavaScript APIs known to be used by fingerprinting scripts, hereafter referred to as ``fingerprinting APIs''.
Fingerprinting APIs include functions and properties that are used in fingerprintjs2 and those discovered by \framework in Section~\ref{sec:new-apis}.
Note that this countermeasure does not at all rely on \framework's detection of fingerprinting scripts.

\item \textbf{Targeted API Restriction.} We restrict access to fingerprinting APIs only for the scripts served from domains that are detected by \framework to deploy fingerprinting scripts.

\item \textbf{Request Blocking.} We block the requests to download the scripts served from domains that are detected by \framework to deploy fingerprinting scripts.

\item \textbf{Hybrid.} We block the requests to download the scripts served from domains that are detected by \framework to deploy fingerprinting scripts, except for first-party and inline scripts.
Additionally, we restrict access to fingerprinting APIs for first-party and inline scripts on detected domains. This protects against active fingerprinting by first parties and both active and passive fingerprinting by third parties.

\end{enumerate}

\section{Evaluation}
\label{sec:evaluation}
We evaluate \framework's performance in terms of its accuracy in detecting fingerprinting scripts and its impact on website breakage when mitigating fingerprinting.

\subsection{Accuracy}
We require samples of fingerprinting and non-fingerprinting scripts to train our supervised machine learning models.
Up-to-date ground truth for fingerprinting is not readily available.
Academic researchers have released lists of scripts \cite{Englehardt16MillionSiteMeasurementCCS,Das18MobileSensorsCCS}, however these only show a snapshot at the time of the paper's publication and are not kept up-to-date.
While many anti-tracking lists (e.g., EasyPrivacy) do include some fingerprinting domains, Disconnect's tracking protection list \cite{disconnect_me} is the only publicly available list that does not lump together different types of tracking and separately identifies fingerprinting domains.
However, Disconnect's list is insufficient for our purposes.
First, Disconnect's list only includes the domain names of companies that deploy fingerprinting scripts, rather than the actual URLs of the fingerprinting scripts.
This prevents us from using the list to differentiate between fingerprinting and non-fingerprinting resources served from those domains.
Second, the list appears to be focused on fingerprinting deployed by popular third-party vendors.
Since first-party fingerprinting is also prevalent \cite{Das18MobileSensorsCCS}, we would like to train our classifier to detect both first- and third-party fingerprinting scripts.
Given the limitations of these options, we choose to detect fingerprinting scripts using a slightly modified version of the heuristics implemented in  \cite{Englehardt16MillionSiteMeasurementCCS}.

\subsubsection{Fingerprinting Definition}
\label{section:fingerprinting-definition}
The research community is not aligned on a single definition to label fingerprinting scripts.
It is often difficult to determine the \textit{intent} behind any individual API access, and classifying all instances of device information collection as fingerprinting will result in a large number of false positives.
For example, an advertisement script may collect a device's screen size to determine whether an ad was viewable and may never use that information as part of a fingerprint to identify the device.
With that in mind, we take a conservative approach: we consider a script as fingerprinting if it uses \texttt{Canvas}, \texttt{WebRTC}, \texttt{Canvas Font}, or \texttt{AudioContext} as defined in \cite{Englehardt16MillionSiteMeasurementCCS}.
Specifically, if the heuristics trigger for any of the above mentioned behaviors, we label the script as fingerprinting and otherwise label it as non-fingerprinting.
We do not consider the collection of attributes from navigator or screen APIs from a script as fingerprinting, as these APIs are frequently used in non-distinct ways by scripts that do not fingerprint users.
We decide to initially use this definition of fingerprinting because it is precise, i.e., it has a low false positive rate.
A low false positive rate is crucial for a reliable ground truth as the classifiers effectiveness will depend on the soundness of ground truth.
The exact details of heuristics are listed in Appendix \ref{app:fp-heuristics}.

\subsubsection{Data Collection}
\label{section:data-collection}
We use our extended version of OpenWPM to crawl the homepages of twenty thousand websites sampled from the Alexa top-100K websites.
To build this sample, we take the top-10K sites from the list and augment it with a random sample of 10K sites with Alexa ranks from 10K to 100K.
This allows us to cover both the most popular websites as well as websites further down the long tail.
During the crawl we allow each site 120 seconds to fully load before timing out the page visit.
We store the HTTP response body content from all documents and scripts loaded on the page as well as the execution traces of all scripts.

Our crawled dataset consists of \SuccesfullWebsitesCrawled websites with \DistinctExecutingScripts distinct executing scripts.
Since we generate our ground truth by analyzing script execution traces, we are only able to collect ground truth from scripts that actually execute during our crawl.
Although we are not able train our classifier on scripts that do not execute during our crawl, we are still able to classify them. Their classification result will depend entirely on the static features extracted from the script contents.
For static features, we successfully create ASTs for \SuccesfullASTs scripts---\FailedASTs scripts (\FailedASTsPercent) fail because of invalid syntax.
Out of valid scripts, we extract a total of \RawStaticFeatures \texttt{parent:child} combinations and do feature selection as described in Section \ref{sec:methodology}.
Specifically, we first filter by a variance threshold of \VarinaceThreshold to reduce the set to \VarinaceThresholdFeatures \texttt{parent:child} combinations.
We then select top 1K features when sorted by information gain.
For dynamic features, we extract a total of \RawDynamicFeatures features from \DistinctExecutingScripts scripts.
Similar to static analysis, we do feature selection as described in Section \ref{sec:methodology} and reduce the feature set to top 1K when sorted by information gain.

\subsubsection{Enhancing Ground Truth}
\label{subsubsection:enhancing-ground-truth}
As discussed in Section \ref{sec:background}, heuristics suffer from two inherent problems.
First, heuristics are narrowly defined which can cause them to miss some fingerprinting scripts.
Second, heuristics are predefined and are thus unable to keep up with evolving fingerprinting scripts.
Due to these problems, we know that our heuristics-based ground truth is imperfect and a machine learning model trained on such a ground truth may perform poorly.
We address these problems by enhancing the ground truth through iterative re-training.
We first train a base model with incomplete ground truth, and then manually analyze the disagreements between the classifier's output and the ground truth.
We update the ground truth whenever we find that our classifier makes a correct decision that was not reflected in the ground truth (i.e., discovers a fingerprinting script that was missed by the ground truth heuristics).
We perform three iterations of this process.

\textbf{Manual labeling.} The manual process of analyzing scripts during iterative re-training works as follows.
We automatically create a report for every script that requires manual analysis.
Each report contains: (1) all of the API method calls and property accesses monitored by our instrumentation, including the arguments and return values, (2) snippets from the script that capture the surrounding context of calls to the APIs used for canvas, WebRTC, canvas font, and AudioContext fingerprinting, (3) a fingerprintjs2 similarity score,\footnote{We compute Jaccard similarity between the script, by first beautifying it and then tokenizing it based on white spaces, and all releases of fingerprintjs2. The release with the highest similarity is reported along with the similarity score.} and (4) the formatted contents of the complete script.
We then manually review the reports based on our domain expertise to determine whether the analyzed script is fingerprinting.
Specifically, we look for heuristic-like behaviors in the scripts.
The heuristic-like behavior means that the fingerprinting code in the script: 

\begin{enumerate}
  \item Is similar to known fingerprinting code in terms of its functionality and structure,
  \item It is accompanied with other fingerprinting code (i.e. most fingerprinting scripts use multiple fingerprinting techniques), and 
  \item It does not interact with the functional code in the script. 
\end{enumerate}

For example, common patterns include sequentially reading values from multiple APIs, storing them in arrays or dictionaries, hashing them, and sending them in a network request without interacting with other parts of the script or page.

\textbf{Findings.} We found the majority of reviews to be straightforward---the scripts in question were often similar to known fingerprinting libraries and they frequently use APIs that are used by other fingerprinting scripts.
If we find any fingerprinting functionality within the script we label the whole script as fingerprinting, otherwise we label it is non-fingerprinting.
To be on the safe side, scripts for which we were unable to make a manual determination (e.g., due to obfuscation) were considered non-fingerprinting.

Overall, perhaps expected, we find that our ground truth based on heuristics is high precision but low recall within the disagreements we analyzed.
Most of the scripts that heuristics detect as fingerprinting do include fingerprinting code, but we also find that the heuristics miss some fingerprinting scripts.
There are two major reasons scripts are missed.
First, the fingerprinting portion of the script resides in a dormant part of the script, waiting to be called by other events or functions in a webpage.
For example, the snippet in Script~\ref{lst:sample_dormant_script} (Appendix \ref{app:deviating-dormant}) defines fingerprinting-specific prototypes and assign them to a \texttt{window} object which can be called at a later point in time.
Second, the fingerprinting functionality of the script deviates from the predefined heuristics.
For example, the snippet in Script~\ref{lst:sample_deviating_script} (Appendix \ref{app:deviating-dormant}) calls \texttt{save} and \texttt{restore} methods on \texttt{CanvasRenderingContext2D} element, which are two method calls used by the heuristics to filter out non-fingerprinting scripts \cite{Englehardt16MillionSiteMeasurementCCS}.

\begin{table}[!t]
    \centering
    \setlength\tabcolsep{1.5pt} 
    \begin{tabular}{l|cc|cccc|cc}
      \toprule
      \multirow{2}{*}{\bfseries Itr.} &
        \multicolumn{2}{c|}{\bfseries Initial} &
        \multicolumn{2}{c}{\bfseries New Detections} &
        \multicolumn{2}{c|}{\bfseries Correct Detections} &
        \multicolumn{2}{c}{\bfseries Enhanced} \\
        & {\bfseries FP} & {\bfseries NON-FP} & {\bfseries FP} & {\bfseries NON-FP} & {\bfseries FP} & {\bfseries NON-FP} & {\bfseries FP} & {\bfseries NON-FP}\\
        \midrule
        \bfseries S1  & 884   & 142,642 & 150 & 232 & 103 & 10 & 977   & 142,549 \\
        \bfseries S2  & 977   & 142,549 & 109 & 182 & 84  & 5  & 1,056 & 142,470 \\
        \bfseries S3  & 1,056 & 142,470 & 76  & 158 & 53  & 1  & 1,108 & 142,418 \\
        \midrule
        \bfseries D1 & 928   & 152,426 & 11  & 52  & 4   & 9  & 923   & 152,431 \\
        \bfseries D2 & 923   & 152,431 & 8   & 35  & 4   & 1  & 926   & 152,428 \\
        \bfseries D3 & 926   & 152,428 & 13  & 36  & 5   & 2  & 929   & 152,425 \\
      \bottomrule
    \end{tabular}
    \caption{Enhancing ground truth with multiple iterations of retaining. Itr. represents the iteration number of training with static (S) and dynamic (D) models. New Detections (FP) represent the additional fingerprinting scripts detected by the classifier and New Detections (NON-FP) represent the new non-fingerprinting scripts detected by the classifier as compared to heuristics. Whereas Correct Detections (FP) represent the manually verified correct determination of the classifier for fingerprinting scripts and Correct Detections (NON-FP) represent the manually verified correct determination of the classifier for non-fingerprinting scripts.}
    \label{table:ground-truth-improvement}
     \vspace{-12pt}
\end{table}

However, for a small number of scripts, the heuristics outperform the classifier.
Scripts which make heavy use of an API used that is used for fingerprinting, and which have limited interaction with the webpage, are sometimes classified incorrectly.
For example, we find cases where the classifier mislabels non-fingerprinting scripts that use the Canvas API to create animations and charts, and which only interact with a few HTML elements in the process.
Since heuristics cannot generalize over fingerprinting behaviors, they do not classify partial API usage and limited interaction as fingerprinting.
In other cases, the classifier labels fingerprinting scripts as non-fingerprinting because they include a single fingerprinting technique along with functional code.
For example, we find cases where classifier mislabels fingerprinting scripts embedded on login pages that only include canvas font fingerprinting alongside functional code.
Since heuristics are precise, they do not consider functional aspects of the scripts and do not classify limited usage of fingerprinting as non-fingerprinting.

\textbf{Improvements.} Table~\ref{table:ground-truth-improvement} presents the results of our manual evaluation for ground truth improvement for both static and dynamic analysis.
It can be seen from the table that our classifier is usually correct when it classifies a script as fingerprinting in disagreement with the ground truth.
We discover new fingerprinting scripts in each iteration.
In addition, it is also evident from the table that our models are able to correct its mistakes with each iteration (i.e., correct previously incorrect non-fingerprinting classifications).
This demonstrates the ability of classifier in iteratively detecting new fingerprinting scripts and correct mistakes as ground truth is improved.
We further argue that this iterative improvement with re-training is essential for an operational deployment of a machine learning classifier and we empirically demonstrate that for \framework.
Overall, we enhance our ground truth by labeling an additional 240 scripts as fingerprinting and 16 scripts as non-fingerprinting for static analysis, as well as 13 scripts as fingerprinting and 12 scripts as non-fingerprinting for dynamic analysis.
In total, we detect 1,108 fingerprinting scripts and 142,418 non-fingerprinting scripts with static analysis and 929 fingerprinting scripts and 152,425 non-fingerprinting scripts using dynamic analysis.

\subsubsection{Classification Accuracy}
\label{sec:results}
We use the decision tree models described in Section~\ref{sec:methodology} to classify the crawled scripts.
To establish confidence in our models against unseen scripts, we perform standard 10-fold cross validation.
We determine the accuracy of our models by comparing the predicted label of scripts with the enhanced ground truth described in Section~\ref{subsubsection:enhancing-ground-truth}.
For the model trained on static features, we achieve an accuracy of 99.8\%, with 85.5\% recall, and 92.7\% precision.
For the model trained on dynamic features, we achieve an accuracy of 99.9\%, with 96.7\% recall, and 99.1\% precision.

\begin{table*}[!htpb]
    \centering
    \begin{tabular}{l|cc|cccccc}
    \toprule
    \bfseries Classifier & \bfseries Heuristics (Scripts/Websites) & \bfseries Classifiers (Scripts/Websites) & \bfseries FPR  & \bfseries FNR & \bfseries Recall & \bfseries Precision & \bfseries Accuracy \\
    \midrule
    \bfseries Static   & 884 / 2,225  & 1,022 / 3,289 & 0.05\%  & 15.7\% & 85.5\% & 92.7\% & 99.8\% \\
    \bfseries Dynamic  & 928 / 2,272 & 907 / 3,278  & 0.005\% & 5.3\%  & 96.7\% & 99.1\% & 99.9\% \\
    \midrule
    \bfseries Combined & 935 / 2,272   & 1,178 / 3,653 & 0.05\%  & 6.1\%  & 93.8\% & 93.1\% & 99.9\% \\
    \bottomrule
\end{tabular}
\caption{\framework's classification results in terms of recall, precision, and accuracy in detecting fingerprinting scripts. ``Heuristics (Scripts/Websites)'' represents the number of scripts and websites detected by heuristics and ``Classifiers (Scripts/Websites)'' represents the number of scripts and websites detected by the classifiers. FPR represents false positive rate and FNR represent false negative rate.}
\label{table:all-results}
 \vspace{-15pt}
\end{table*}

\textbf{Combining static and dynamic models.}
In \framework, we train two separate machine learning models---one using features extracted from the static representation of the scripts, and one using features extracted from the dynamic representation of the scripts.
Both of the models provide complementary information for detecting fingerprinting scripts.
Specifically, the model trained on static features identifies dormant scripts that are not captured by the dynamic representation, whereas the model trained on dynamic features identifies obfuscated scripts that are missed by the static representation.
We achieve the best of both worlds by combining the classification results of these models.
We combine the models by doing an \texttt{OR} operation on the results of each model.
Specifically, if either of the model detects a script as fingerprinting, we consider it a fingerprinting script.
If neither of the model detects a script as fingerprinting, then we consider it a non-fingerprinting script.
We manually analyze the differences in detection of static and dynamic models and find that the \DormantDetectedByStatic of scripts identified only by the static model are partially or completely dormant and \ObfuscatedDetectedByDynamic of the scripts identified only by the dynamic model are obfuscated or excessively minified. 

Table~\ref{table:all-results} presents the combined and individual results of static and dynamic models.
It can be seen from the table that \framework's classifier detects \PercentageMoreDetectionByClassifiers more scripts than the heuristics with a negligible false positive rate (FPR) of \ClassifierFPR and a false negative rate (FNR) of \ClassifierFNR.
Overall, we find that by combining the models, \framework increases its detection rate by almost 10\% and achieves an overall accuracy of 99.9\% with 93.8\% recall and 93.1\% precision.\footnote{
Is the complexity of a machine learning model really necessary? Would a simpler approach work as well? 
While our machine learning model performs well, we seek to answer this question in Appendix~\ref{sec:why-ml} by comparing our performance to a more straightforward similarity approach to detect fingerprinting.
We compute the similarity between scripts and the popular fingerprinting library fingerprintjs2.
Overall, we find that script similarity not only detects a partial number of fingerprinting scripts detected by our machine learning model but also incurs an unacceptably high number of false positives.
}

\subsection{Breakage}
\label{sec:enforcement}

We implement the countermeasures listed in Section \ref{subsec: mitigating} in a browser extension to evaluate their breakage. 
The browser extension contains the countermeasures as options that can be selected one at a time.
For API restriction, we override functions and properties of fingerprinting APIs and return an error message when they are accessed on any webpage.
For targeted API restriction, we extract a script's domain by traversing the stack each time the script makes a call to one of the fingerprinting APIs.
We use \framework's classifier determinations to create a domain-level (eTLD+1, which matches Disconnect's fingerprinting list used by Firefox) filter list.
For request blocking, we use the \texttt{webRequest} API \cite{webrequest_api} to intercept and block outgoing web requests that match our filter list \cite{CliqzContentBlockingLibrary}.

Next, we analyze the breakage caused by these enforcements on a random sample of 50 websites that load fingerprinting scripts along with 11 websites that are reported as broken in Firefox due to fingerprinting countermeasures \cite{firefox_fp_block_breakage}.
Prior research \cite{Iqbal20AdGraphSP,snyder2017most} has mostly relied on manual analysis to analyze website breakage due the challenges in automating breakage detection.
We follow the same principles and manually analyze website breakage under the four fingerprinting countermeasures.
To systemize manual breakage analysis, we create a taxonomy of common fingerprinting breakage patterns by going through the breakage-related bug reports on Mozilla's bug tracker \cite{firefox_fp_block_breakage}.
We open each test website on vanilla Firefox (i.e., without our extension installed) as control and also with our extension installed as treatment.
It is noteworthy that we disable Firefox's default privacy protections in both the control and treatment branches of our study to isolate the impact of our protections.
We test each of the countermeasures one by one by trying to interact with the website for few minutes by scrolling through the page and using the obvious website functionality.
If we discover missing content or broken website features only in the treatment group, we assign a breakage label using the following taxonomy:

\begin{enumerate}
        \item \textbf{Major:} The core functionality of the website is broken. Examples include: login or registration flow, search bar, menu, and page navigation.
        \item \textbf{Minor:} The secondary functionality of the website is broken. Examples include: comment sections, reviews, social media widgets, and icons.
        \item \textbf{None:} The core and secondary functionalities of the website are the same in treatment and control. We consider missing ads as no breakage.
\end{enumerate}

\begin{table}[!ht]
    \centering
    \begin{tabular}{l|cc|c}
    \toprule
    \bfseries Policy  &  \bfseries Major (\%)  & \bfseries  Minor (\%) & \bfseries Total (\%)\\
    \hline
    Blanket API restriction & 48.36\%   & 19.67\% &  \cellcolor{MyMaroon-f60000}68.03\% \\
    \hline
    Targeted API restriction & 24.59\%   & 5.73\%  &  \cellcolor{MyMaroon-ffbbbb}30.32\% \\
    \hline
    Request blocking   & 44.26\%   & 5.73\%  &  \cellcolor{MyMaroon-ff3232}50\% \\
    \hline
    Hybrid  & 38.52\%   & 8.19\%  &  \cellcolor{MyMaroon-fc6565}46.72\% \\
    \bottomrule
\end{tabular}
\caption{Breakdown of breakage caused by different countermeasures. The results present the average assessment of two reviewers.}
\label{table:breakage-analysis}
\end{table}

To reduce coder bias and subjectivity, we asked two reviewers to code the breakage on the full set of 61 test websites using the aforementioned guidelines.
The inter-coder reliability between our two reviewers is 87.70\% for a total of 244 instances (4 countermeasures $\times$ 61 websites).
Table \ref{table:breakage-analysis} summarizes the averaged breakage results.
Overall, we note that targeted countermeasures that use \framework's detection reduce breakage by a factor of 2 on the tested websites that are particularly prone to breakage.\footnote{These websites employ fingerprinting scripts and/or are reported to be broken due to fingerprinting-specific countermeasures. Thus, they represent a particularly challenging set of websites to evaluate breakage by fingerprinting countermeasures.}
More specifically, blanket API restriction suffers the most (breaking more than two-thirds of the tested websites) while the targeted API restriction causes the least breakage (with no major breakage on about 75\% of the tested websites).

Surprisingly, we find that the blanket API restriction causes more breakage than request blocking.
We posit this is caused by the fact that blanket API restriction is applied to all scripts on the page, regardless of whether they are fingerprinting, since even benign functionality may be impacted.
By comparison, request blocking only impacts scripts known to fingerprint.
Next, we observe that targeted API restrictions has the least breakage.
%
This is expected, as we do not block requests and only limit scripts that are suspected of fingerprinting; the functionality of benign scripts is not impacted.

We find that the hybrid countermeasure causes less breakage than request blocking but more breakage than the targeted API restrictions.
The hybrid countermeasure performs better than request blocking because it does not block network requests to load first-party fingerprinting resources and instead applies targeted API restrictions to protect against first-party fingerprinting.
Whereas it performs worse than targeted API restrictions because it still blocks network requests to load third-party fingerprinting resources that are not blocked by the targeted API restrictions.
Though hybrid blocking causes more breakage than targeted API restriction, it offers the best protection.
Hybrid blocking mitigates both active and passive fingerprinting from third-party resources, and active fingerprinting from first-party resources and inline scripts.
The only thing missed by hybrid blocking---passive first-party fingerprinting---is nearly impossible to block without breaking websites because any first-party resource loaded by the browser can passively collect device information.

We find that the most common reason for website breakage is the dependence of essential functionality on fingerprinting code. 
In severe cases, registration/login or other core functionality on a website depends on computing the fingerprint.
For example, the registration page on \url{freelancer.com} is blank because we restrict the fingerprinting script from \url{f-cdn.com}.
In less severe cases, websites embed widgets or ads that rely on fingerprinting code.
For example, the social media widgets on \url{ucoz.ru/all/} disappears because we apply restrictions to the fingerprinting script from \url{usocial.pro}.

\section{Measuring Fingerprinting In The Wild}
\label{sec:in-the-wild}
Next, we use the detection component of \framework to analyze the state of fingerprinting on top-100K websites.
To collect data from the Alexa top-100K websites, we first start with the 20K website crawl described in Section~\ref{section:data-collection}, and follow the same collection procedure for the remaining 80K websites not included in that measurement.
Out of this additional 80K, we successfully visit 71,112 websites.
The results provide an updated view of fingerprinting deployment following the large-scale 2016 study by Englehardt and Narayanan \cite{Englehardt16MillionSiteMeasurementCCS}.
On a high-level we find: (1) the deployment of fingerprinting is still growing---reaching over a quarter of the Alexa top-10K sites, (2) fingerprinting is almost twice as prevalent on news sites than in any other category of site, (3) fingerprinting is used for both anti-ad fraud and potential cross-site tracking.

\subsection{Over a quarter of the top sites now fingerprint users}
\label{section:prevalence-top-ranking}
We first examine the deployment of fingerprinting across the top sites; our results are summarized in Table~\ref{table:alexa-fp-range}.
In alignment with prior work \cite{Englehardt16MillionSiteMeasurementCCS}, we find that fingerprinting is more prevalent on highly ranked sites.
We also detect more fingerprinting than prior work \cite{Englehardt16MillionSiteMeasurementCCS}, with over a quarter of the top sites now deploying fingerprinting.
This increase in use holds true across all site ranks---we observe a notable increase even within less popular sites (i.e., 10K - 100K).
Overall, we find that more than \TopTenKFingerprintingWebsites of top-100K websites deploy fingerprinting.

We also find significantly more domains serving fingerprinting than past work---2,349 domains on the top 100K sites (Table~\ref{table:top-fp-vendors}) compared to 519 domains\footnote{Englehardt and Narayanan \cite{Englehardt16MillionSiteMeasurementCCS} do not give an exact count of the number of domains serving fingerprinting across all measured techniques, and instead give a count for each individual fingerprinting technique. To get an upper bound on the total count, we assume there is no overlap between the reported results of each technique and take the sum.} on the top 1 million sites \cite{Englehardt16MillionSiteMeasurementCCS}.
This suggests two things: our method is detecting a more comprehensive set of techniques than measured by Englehardt and Narayanan \cite{Englehardt16MillionSiteMeasurementCCS}, and/or that the use of fingerprinting---both in prevalence and in the number of parties involved---has significantly increased between 2016 and 2019.

\begin{table}[h]
    \centering
    \begin{tabular}{l|rr}
    \toprule
    \bfseries Rank Interval & \bfseries Websites (count) & \bfseries Websites (\%)\\
    \midrule
    1 to 1K     & 266 & 30.60\% \\
    1K to 10K   & 2,010 & 24.45\% \\
    10K to 20K  & 981 & 11.10\% \\
    20K to 50K  & 2,378 & 8.92\% \\
    50K to 100K & 3,405 & 7.70\% \\
    \midrule
    1 to 100K   & 9,040 & 10.18\% \\
    \bottomrule
\end{tabular}
\caption{Distribution of Alexa top-100K websites that deploy fingerprinting. Results are sliced by site rank.}
\label{table:alexa-fp-range}
\end{table}

\subsection{Fingerprinting is most common on news sites}
\begin{figure}[h]
    \centering
    \includegraphics[width=0.45\textwidth]{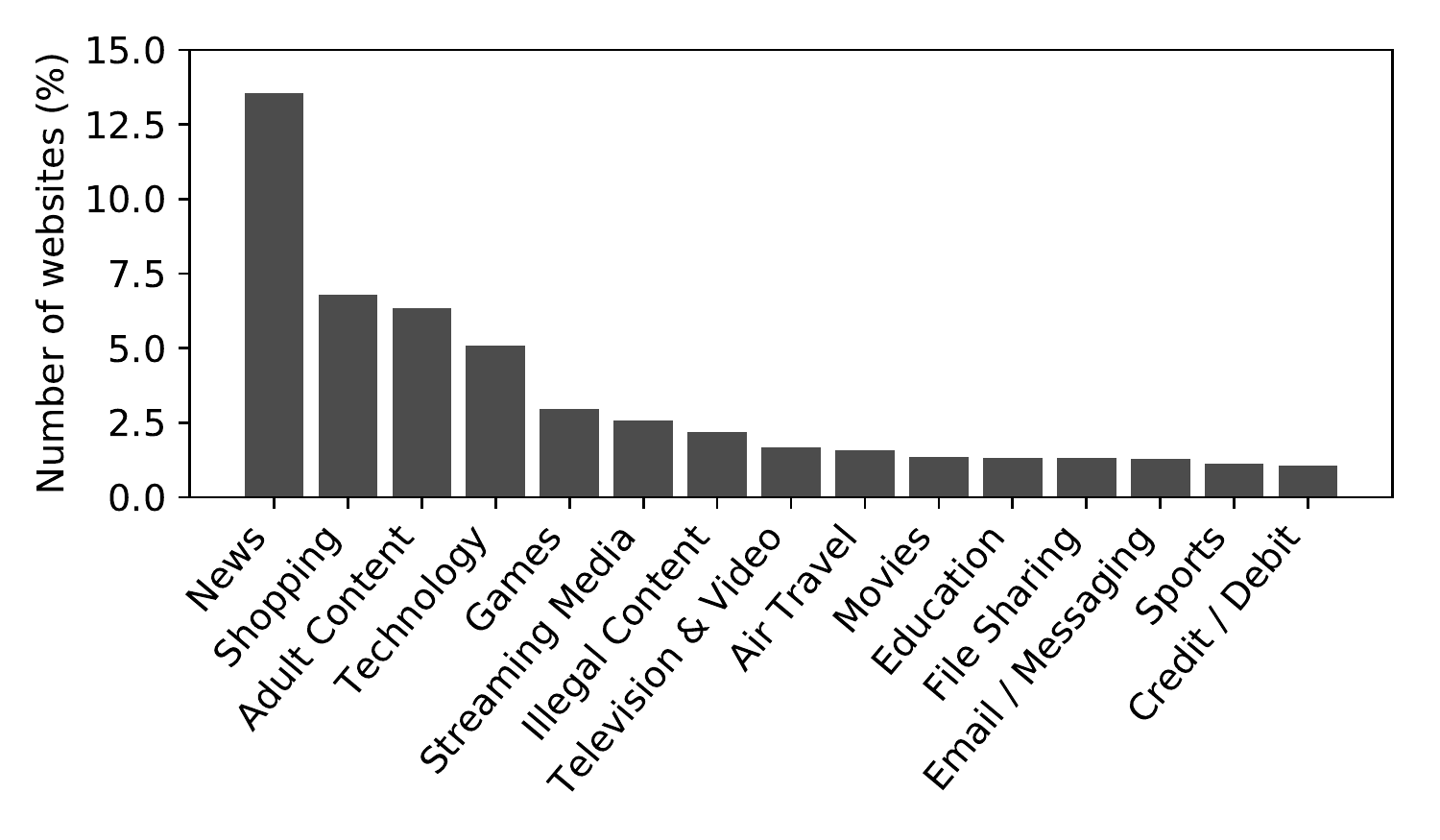}
     \vspace{-10pt}
    \caption{The deployment of fingerprinting scripts across different categories of websites.}
\label{figure:fp-website-categories}
 \vspace{-20pt}
\end{figure}

Fingerprinting is deployed unevenly across different categories of sites.\footnote{We use Webshrinker \cite{webshrinker_website} for website categorization API.}
The difference is staggering---ranging from nearly 14\% of news websites to just 1\% of credit/debit related websites.
Figure~\ref{figure:fp-website-categories} summarizes our findings.

The distribution of fingerprinting scripts in Figure~\ref{figure:fp-website-categories} roughly matches the distribution of trackers (i.e., not only fingerprinting, but any type of tracking) measured in past work \cite{Englehardt16MillionSiteMeasurementCCS}.
One possible explanation of these results is that---like traditional tracking methods---fingerprinting is more common on websites that rely on advertising for monetization.
Our results in Section~\ref{sec:fp-use-cases} reinforce this interpretation, as the most prevalent vendors classified as fingerprinting provide anti-ad fraud and tracking services.
The particularly high use of fingerprinting on news websites could also point to fingerprinting being used as part of paywall enforcement, since cookie-based paywalls are relatively easy to circumvent \cite{Papadopoulos20paywallWWW}.

\subsection{Fingerprinting is used to fight ad fraud but also for potential cross-site tracking}
\label{sec:fp-use-cases}
Fingerprinting scripts detected by \framework are often served by third-party vendors.
Three of the top five vendors in Table~\ref{table:top-fp-vendors} (\url{doubleverify.com}, \url{adsafeprotected.com}, and \url{adsco.re}) specialize in verifying the authenticity of ad impressions.
Their privacy policies mention that they use ``device identification technology'' that leverages ``browser type, version, and capabilities'' \cite{doubleverify_privacy_policy,Integral_ad_science_privacy_policy,adscore_privacy_policy}.
Our results also corroborate that bot detection services rely on fingerprinting \cite{Azad20WebRunnerDIMVA}, and indicate that prevalent fingerprinting vendors provide anti-ad fraud services.
The two remaining vendors of the top five, i.e., \url{alicdn.com} and \url{yimg.com}, appear to be CDNs for Alibaba and Oath/Yahoo!, respectively.

\begin{table}[h]
    \centering
    \begin{tabular}{l|cr}
    \toprule
    \bfseries Vendor Domain & \bfseries Tracker & \bfseries Websites (count) \\
    \midrule
    doubleverify.com    & Y             & 2,130 \\
    adsafeprotected.com & Y             & 1,363\\
    alicdn.com          & N             & 523\\
    adsco.re            & N             & 395\\
    yimg.com            & Y             & 246\\
    2,344 others        & Y(86)         & 5,702\\
    \midrule
    Total               &               & 10,359 (9,040 distinct)\\
    \bottomrule
\end{tabular}
\caption{The presence of the top vendors classified as fingerprinting on Alexa top-100K websites. Tracker column shows whether the vendor is a cross-site tracker according to Disconnect's tracking protection list. Y represents yes and N represents no.}
\label{table:top-fp-vendors}
\end{table}

Several fingerprinting vendors disclose using cookies ``to collect information about advertising impression opportunities'' \cite{Integral_ad_science_privacy_policy} that is shared with ``customers and partners to perform and deliver the advertising and traffic measurement services'' \cite{doubleverify_privacy_policy}.
To better understand whether these vendors participate in cross-site tracking, we first analyze the overlap of the fingerprinting vendors with Disconnect's tracking protection list \cite{disconnect_me}.\footnote{We exclude the cryptomining and fingerprinting categories of the Disconnect list. The list was retrieved in June 2019.}
Disconnect employs a careful manual review process \cite{disconnect_tracking_definition} to classify a service as tracking.
For example, Disconnect classifies c3tag as tracking \cite{c3metrics_policy_review_disconnect,c3metrics_privacy_policy} and adsco.re as not tracking \cite{adscore_policy_review_disconnect,adscore_privacy_policy} because, based on their privacy policies, the former shares Personally Identifiable Information (PII) with its customers while the latter does not. 
We find that 3.78\% of the fingerprinting vendors are classified as tracking by Disconnect.

We also analyze whether fingerprinting vendors engage in cookie syncing \cite{Papadopoulos19CookieWWW}, which is a common practice by online advertisers and trackers to improve their coverage.
For example, a tracker may associate browsing data from a single device to multiple distinct identifier cookies when cookies are cleared or partitioned.
However, a fingerprinting vendor can use a device fingerprint to link those cookie identifiers together \cite{Englehardt14HiddenPerils}.
If the fingerprinting vendor had previously cookie synced with other trackers, it can use its fingerprint to link cookies for other trackers.
We use the list by Fouad et al. \cite{Fouad20PixelPETS} to identify fingerprinting domains that also participate in cookie syncing.
We find that 17.28\% of the fingerprinting vendors participate in cookie syncing.
More importantly, we find that fingerprinting vendors often sync cookies with well-known ad-tech vendors.
For example, \url{adsafeprotected.com} engages in cookie syncing with \url{rubiconproject.com} and \url{adnxs.com}.
We also find that many fingerprinting vendors engage in cookie syncing with numerous third-parties. 
For example, \url{openx.net} engages in cookie syncing with 332 other domains, out of which 14 are classified as tracking by Disconnect.
We leave an in-depth large-scale investigation of the interplay between fingerprinting and cookie syncing as future work.

\section{Analyzing APIs used by Fingerprinters}
\label{sec:new-apis}
In this section, we are interested in systematically investigating whether any newly proposed or existing JavaScript APIs are being exploited for browser fingerprinting.
There are serious concerns that newly proposed or existing JavaScript APIs can be exploited in unexpected ways for browser fingerprinting \cite{w3cFPGuidance}.

We start off by analyzing the distribution of Javascript APIs in fingerprinting scripts.
Specifically, we extract Javascript API keywords (i.e., API names, properties, and methods) from the source code of scripts and sort them based on the ratio of their fraction of occurrence in fingerprinting scripts to the fraction of occurrence in non-fingerprinting scripts.
This ratio captures the relative prevalence of API keywords in fingerprinting scripts as compared to non-fingerprinting scripts. 
A higher value of the ratio for a keyword means that it is more prevalent in fingerprinting scripts than non-fingerprinting scripts. 
Note that $\infty$ means that the keyword is only present in fingerprinting scripts. 
Table \ref{table:fp-api-usage} lists some of the interesting API keywords that are disproportionately prevalent in fingerprinting scripts. 
We note that some APIs are primarily used by fingerprinting scripts, including APIs which have been reported by prior fingerprinting studies (e.g., \texttt{accelerometer}) and those which have not (e.g., \texttt{getDevices}).
We present a more comprehensive list of the API keywords disproportionately prevalent in fingerprinting scripts in Appendix \ref{app:high-frequency-fp-apis}.

\begin{table}[htpb]
    \centering
    \begin{tabular}{l|rrr}
    \toprule
    \bfseries Keywords & \bfseries Ratio & \bfseries Scripts (count) & \bfseries Websites (count) \\
    \midrule
    MediaDeviceInfo & $\infty$ & 1 & 1363\\
    magnetometer & $\infty$ & 215 & 241\\
    PresentationRequest & $\infty$ & 16 & 16\\
    onuserproximity & 543.77 & 18 & 18\\
    accelerometer & 326.71 & 219 & 247\\
    chargingchange & 302.10 & 20 & 20\\
    getDevices & 187.62 & 59 & 80\\
    maxChannelCount & 184.44 & 29 & 40\\
    baseLatency & 181.26 & 3 & 8\\
    vibrate & 57.68 & 232 & 1793\\
    \bottomrule
\end{tabular}
\caption{A sample of frequently used JavaScript API keywords in fingerprinting scripts and their presence on 20K websites crawl. Scripts (count) represents the number of distinct fingerprinting scripts in which the keyword is used and Websites (count) represents the number of websites on which those scripts are embedded.}
\label{table:fp-api-usage}
\vspace{-10pt}
\end{table}

Since the number of API keywords is quite large, it is practically infeasible to manually analyze all of them.
Thus, we first group the extracted API keywords into a few clusters and then manually analyze the cluster which has the largest concentration of API keywords that are disproportionately used in the fingerprinting scripts detected by \framework.
Our key insight is that browser fingerprinting scripts typically do not use a technique (e.g., canvas fingerprinting) in isolation but rather combine several techniques together.
Thus, we expect fingerprinting-related API keywords to separate out as a distinct cluster.

To group API keywords into clusters, we first construct the co-occurrence graph of API keywords.
Specifically, we model API keywords as nodes and include an edge between them that is weighted based on the frequency of co-occurrence in a script.
Thus, co-occurring API keywords appear together in our graph representation.
We then partition the API keyword co-occurrence graph into clusters by identifying strongly connected communities of co-occurring API keywords.
Specifically, we extract communities of co-occurring keywords by computing the partition of the nodes that maximize the modularity using the Louvain method \cite{communityBlondel2008}.
In total, we extract 25 clusters with noticeable dense cliques of co-occurring API keywords.
To identify the clusters of interest, we assign the API keyword's fraction of occurrence in fingerprinting scripts to the fraction of occurrence in non-fingerprinting scripts as weights to the nodes.
We further classify nodes based on whether they appear in fingerprintjs2 \cite{Fingerprint2js}, which is a popular open-source browser fingerprinting library.

We investigate the cluster with the highest concentration of nodes that tend to appear in the detected fingerprinting scripts and those that appear in fingerprintjs2.
While we discover a number of previously unknown uses of JavaScript APIs by fingerprinting scripts, for the sake of concise discussion, instead of individually listing all of the previously unknown JavaScript API keywords, we thematically group them.
We discuss how each new API we discover to be used by fingerprinting scripts may be abused to extract identifying information about the user or their device.
While our method highlights \textit{potential} abuses, a deep manual analysis of each script is required to confirm abuse.

\textbf{Functionality fingerprinting.}
This category covers browser fingerprinting techniques that probe for different functionalities supported by the browser.
Modern websites rely on many APIs to support their rich functionality.
However, not all browsers support every API or may have the requisite user permission.
Thus, websites may need to probe for APIs and permissions to adapt their functionality.
However, such feature probing can potentially leak entropy.

\begin{enumerate}[wide, labelwidth=!, labelindent=0pt]
\item \textit{Permission fingerprinting:}
    \texttt{Permissions} API provides a way to determine whether a permission is granted or denied to access a feature or an API.
    We discover several cases in which the \texttt{Permissions} API was used in fingerprinting scripts.
    Specifically, we found cases where the status and permissions for APIs such as \texttt{Notification}, \texttt{Geolocation}, and \texttt{Camera} were probed.
    The differences in permissions across browsers and user settings can be used as part of a fingerprint.

\item \textit{Peripheral fingerprinting:}
    Modern browsers provide interfaces to communicate with external peripherals connected with the device.
    We find several cases in which peripherals such as gamepads and virtual reality devices were probed.
    In one of the examples of peripherals probing, we find a case in which keyboard layout was probed using \texttt{getLayoutMap} function.
    The layout of the keyboard (e.g., size, presence of specific keys, string associated with specific keys) varies across different vendors and models.
    The presence and the various functionalities supported by these peripherals can potentially leak entropy.

\item \textit{API fingerprinting:}
    All browsers expose differing sets of features and APIs to the web.
    Furthermore, some browser extensions override native JavaScript methods.
    Such implementation inconsistencies in browsers and modifications by user-installed extensions can potentially leak entropy \cite{SchwarzNDSSJSTemplate2019}.
    We find several cases in which certain functions such as \texttt{AudioWorklet} were probed by fingerprinting scripts.
    \texttt{AudioWorklet} is only implemented in Chromium-based browsers (e.g., Chrome or Opera) starting version 66 and its presence can be probed to check the browser and its version.
    We also find several cases where fingerprinting scripts check whether certain functions such as \texttt{setTimeout} and \texttt{mozRTCSessionDescription} were overridden.
    Function overriding can also leak presence of certain browser extensions.
    For example, Privacy Badger \cite{privacybadger_web} overrides several prototypes of functions that are known to be used for fingerprinting.

\end{enumerate}

\textbf{Algorithmic fingerprinting.}
This category covers browser fingerprinting techniques that do not just simply probe for different functionalities.
These browser fingerprinting techniques algorithmically process certain inputs using different JavaScript APIs and exploit the fact that different implementations process these inputs differently to leak entropy.
We discuss both newly discovered uses of JavaScript APIs that were previously not observed in fingerprinting scripts \emph{and} known fingerprinting techniques that seem to have evolved since their initial discovery.

\begin{enumerate}[wide, labelwidth=!, labelindent=0pt]
\item \textit{Timing fingerprinting:}
    The \texttt{Performance} API provides high-resolution timestamps of various points during the life cycle of loaded resources and it can be used in various ways to conduct timing related fingerprinting attacks \cite{Rols2018ClockCCS, ResourceTiming}.
    We find several instances of fingerprinting scripts using the \texttt{Performance} API to record timing of all its events such as \texttt{domainLookupStart}, \texttt{domainLookupEnd}, \texttt{domInteractive}, and \texttt{msFirstPaint}.
    Such measurements can be used to compute the DNS lookup time of a domain, the time to interactive DOM, and the time of first paint.
    A small DNS lookup time may reveal that the URL has previously been visited and thus can leak the navigation history \cite{ResourceTiming}, whereas time to interactive DOM and time to first paint for a website may vary across different browsers and different underlying hardware configurations.
    Such differences in timing information can potentially leak entropy.

\item \textit{Animation fingerprinting:}
    Similar to timing fingerprinting, we found fingerprinting scripts using \texttt{requestAnimationFrame} to compute the frame rate of content rendering in a browser.
    The browser guarantees that it will execute the callback function passed to \texttt{requestAnimationFrame} before it repaints the view.
    The browser callback rate generally matches the display refresh rate \cite{requestAnimationFrame_api} and the number of callbacks within an interval can capture the frame rate.
    The differences in frame rates can potentially leak entropy.

\item \textit{Audio fingerprinting:}
    Englehardt and Narayanan \cite{Englehardt16MillionSiteMeasurementCCS} first reported the audio fingerprinting technique that uses the \texttt{AudioContext} API.
    Specifically, the audio signal generated with \texttt{AudioContext} varies across devices and browsers.
    Audio fingerprinting seems to have evolved.
    We identify several cases in which fingerprinting scripts used the \texttt{AudioContext} API to capture additional properties such as \texttt{numberOfInputs}, \texttt{numberOfOutputs}, and \texttt{destination} among many others properties.
    In addition to reading \texttt{AudioContext} properties, we also find cases in which \texttt{canPlayType} is used to extract the audio codecs supported by the device.
    This additional information exposed by the \texttt{AudioContext} API can potentially leak entropy.

\item \textit{Sensors fingerprinting:}
    Prior work has shown that the device sensors can be abused for browser fingerprinting \cite{Das18MobileSensorsCCS, Dey14AccelPrintNDSS, BojinovSensorsarXiv2014}.
    We find several instances of previously known and unknown sensors being used by fingerprinting scripts.
    Specifically, we find previously known sensors \cite{Das18MobileSensorsCCS} such as \texttt{devicemotion} and \texttt{deviceorientation} and, more importantly, previously unknown sensors such as  \texttt{userproximity} being used by fingerprinting scripts.
\end{enumerate}

\section{Limitations}
\label{sec:limitations}

In this section, we discuss some of the limitations of \framework's detection and mitigation components. 
Since \framework detects fingerprinting at the granularity of a script, an adversarial website can disperse fingerprinting scripts into several chunks to avoid detection or amalgamate all scripts---functional and fingerprinting---into one to avoid enforcement of mitigation countermeasures.

\textbf{Evading detection through script dispersion.}
For detection, \framework only considers syntactic and semantic relationship within scripts and does not considers relationship across scripts.
Because of its current design, \framework may be challenged in detecting fingerprinting when the responsible code is divided across several scripts. 
However, \framework can be extended to capture interaction among scripts by more deeply instrumenting the browser.
For example, prior approaches such as AdGraph \cite{Iqbal20AdGraphSP} and JSGraph \cite{LiNDSS18JSgraph} instrument browsers to capture cross-script interaction.
Future versions of \framework can also implement such instrumentation; in particular, \framework can be extended to capture the parent-child relationships of script inclusion.
To avoid trivial detection through parent-child relationships, the script dispersion technique would need to be embed each chunk into a website from an independent ancestor node, and return the results to seemingly independent servers. 
Thus, script dispersion also has a maintenance cost: each update to the fingerprinting script will require the distribution of script into several chunks along with extensive testing to ensure correct implementation.

\textbf{Evading countermeasures through script amalgamation.}
To restrict fingerprinting, \framework's most effective countermeasure (i.e. targeted API restriction) is applied at the granularity of a script. 
 \framework may break websites where all of the scripts are amalgamated in a single script. 
However, more granular enforcement can be used to effectively prevent fingerprinting in such cases.
For example, the instrumentation used by future versions of \framework can be extended to track the execution of callbacks and target those related to fingerprinting. 
It is noteworthy that---similar to script dispersion---script amalgamation has a maintenance cost: each update to any of the script will require the amalgamation of all scripts into one.
Script amalgamation could also be used as a countermeasure against ad and tracker blockers, which would introduce the same type of breakage.
However, anecdotal evidence suggests that the barriers to use are sufficiently high to prevent widespread deployment of amalgamation as a countermeasure against privacy tools. 

\section{Conclusion}
\label{sec:conclusion}
We presented \framework, a machine learning based syntactic-semantic approach to accurately detect browser fingerprinting behaviors.
\framework outperforms heuristics from prior work by detecting 26\% more fingerprinting scripts and helps reduce website breakage by 2X.
\framework's deployment showed that browser fingerprinting is more prevalent on the web now than ever before.
Our measurement study on the Alexa top-100K websites showed that fingerprinting scripts are deployed on 10.18\% of the websites by 2,349 different domains.

We plan to report the domains serving fingerprinting scripts to tracking protection lists such as Disconnect \cite{disconnect_me} and EasyPrivacy \cite{EasyPrivacy}.
\framework also helped uncover exploitation of several new APIs that were previously not known to be used for browser fingerprinting.
We plan to report the names and statistics of these APIs to privacy-oriented browser vendors and standards bodies.
To foster follow-up research, we will release our patch to OpenWPM, fingerprinting countermeasures prototype extension, list of newly discovered fingerprinting vendors, and bug reports submitted to tracking protection lists, browser vendors, and standards bodies at \url{https://uiowa-irl.github.io/FP-Inspector}.
\section*{Acknowledgements}
\label{sec:acknowledgement}
The authors would like to thank Charlie Wolfe (NSF REU Scholar) for his help with the breakage analysis.
A part of this work was carried out during the internship of the lead author at Mozilla.
This work is supported in part by the National Science Foundation under grant numbers 1715152, 1750175, 1815131, and 1954224.

{\normalsize \bibliographystyle{acm}}
\bibliography{references}


\section{Appendix}
\label{sec:appendix}

\subsection{Extensions to OpenWPM JavaScript instrumentation}
\label{app:js-extensions}

OpenWPM's instrumentation does not cover a number of APIs used for fingerprinting by prominent libraries---including the Web Graphics Library (WebGL) and \texttt{performance.now}.
These APIs have been discovered to be fingerprintable \cite{laperdrix2019torfingerprinting}.
The standard use case of WebGL is to render 2D and 3D graphics in HTML canvas element, however, it has potential to be abused for browser fingerprinting.
The WebGL renderer and vendor varies by the OS and it creates near distinct WebGL images with same configurations on different machines.
The WebgGL properties and the rendered image are used by current state-of-the-art browser fingerprinting \cite{Fingerprint2js, mathtag} scripts.
Since WebGL is used by popular fingerprinting scripts, we instrument WebGL JavaScript API.
\texttt{performance.now} is another JavaScript API method whose standard use case is to return time in floating point milliseconds since the start of a page load but it also have fingerprinting potential.
Specifically, the timing information extracted from \texttt{performance.now} can be used for timing specific fingerprint attacks such as typing cadence \cite{TORFPBug, TypingFrequencyFingerprinting}.
We extend OpenWPM to also capture execution of \texttt{performance.now}.

For completeness, we instrument additional un-instrumented methods of already instrumented JavaScript APIs in OpenWPM.
Specifically, we enhance our execution trace by instrumenting methods such as \texttt{drawImage} and \texttt{sendBeacon} for \texttt{canvas} and \texttt{navigation} JavaScript APIs, respectively.

Since most fingerprinting scripts use JavaScript APIs that are also used by gaming and interactive websites  (e.g. \texttt{canvas}), we instrument additional JavaScript APIs to capture script's interaction with DOM.
Specifically, to capture DOM interaction specific JavaScript APIs, we instrument \texttt{document}, \texttt{node}, and \texttt{animation} APIs.
JavaScript is an event driven language and it has capability to execute code when events trigger.
To extend our execution trace, we instrument JavaScript events such as \texttt{onmousemove} and \texttt{touchstart} to capture user specific interactions.

In addition, we notice that some scripts make multiple calls to JavaScript API methods such as \texttt{createElement} and \texttt{setAttribute} during their execution.
We limit our recording to only first 50 calls of each method per script, except for \texttt{CanvasRenderingContext2D.measureText} and \texttt{CanvasRenderingContext2D.font}, which are called multiple times for canvas font fingerprinting.
Furthermore, the event driven nature of JavaScript makes it challenging to capture the complete execution trace of scripts.
To this end, to get a comprehensive execution of a script, we synthetically simulate user activity on a webpage.
First, we scroll the wbepage from top to bottom and do random mouse movements to trigger events.
Second, we record all of the events (e.g. \texttt{onscroll}) as they are registered on different elements on a webpage and execute them after 10 seconds of a page load.
Doing so, we synthetically simulate events and capture JavaScript API methods that were waiting for those events to trigger.

\subsection{Sample Features Extracted From ASTs \& Execution Traces}
\label{app:sample-features}
Table \ref{table:static_features_list} shows a sample of the features extracted from the AST in Figure \ref{figure:example_sub_ast} and Table \ref{table:dynamic_features_list} shows a sample of the dynamic features extracted from execution trace of Script \ref{figure:sample_fp_script_obfuscated}.
\vspace{-5pt}
\begin{table}[h]
    \centering
    \begin{tabular}{lr}
        \toprule
        \bfseries Static Features \\
        \midrule
            ArrayExpression:monospace \\
            MemberExpression:font \\
            ForStatement:var \\
            MemberExpression:measureText \\
            MemberExpression:width \\
            MemberExpression:length \\
            MemberExpression:getContext \\
            CallExpression:canvas \\
        \bottomrule
        \\
\end{tabular}
\caption{A sample of features extracted from AST in Figure \ref{figure:example_sub_ast}.}
\label{table:static_features_list}
\end{table}

\begin{table}[h]
    \centering
    \begin{tabular}{l|r}
        \toprule
        \bfseries  Feature Name & \bfseries Feature Value \\
        \midrule
        Document.createElement & True \\
        HTMLCanvasElement.width & True \\
        HTMLCanvasElement.height & True \\
        HTMLCanvasElement.getContext & True \\
        CanvasRenderingContext2D.measureText & True \\
        Element Tag Name & Canvas \\
        HTMLCanvasElement.width & 100 \\
        HTMLCanvasElement.height & 100 \\
        CanvasRenderingContext2D.measureText & 7 (no. of chars.)\\
        CanvasRenderingContext2D.measureText & N (no. of calls)\\
        \bottomrule
\end{tabular}
\caption{A sample of the dynamic features extracted from the execution trace of Script \ref{figure:sample_fp_script_obfuscated}.}
\label{table:dynamic_features_list}
\end{table}

\subsection{Fingerprinting Heuristics}
\label{app:fp-heuristics}
Below we list down the slightly modified versions of heuristics proposed by Englehardt and Narayanan \cite{Englehardt16MillionSiteMeasurementCCS} to detect fingerprinting scripts.
Since non-fingerprinting adoption of fingerprinting APIs have increased since the study, we make modifications to the heuristics to reduce the false positives.
These heuristics are used to build our initial ground truth of fingerprinting and non-fingerprinting scripts.

\point{Canvas Fingerprinting}
A script is identified as canvas fingerprinting script according to the following rules:
    \begin{enumerate}[leftmargin=0pt,itemindent=2em]
    \item The canvas element text is written with \texttt{fillText} or \texttt{strokeText} and style is applied with \texttt{fillStyle} or \texttt{strokeStyle} methods of the rendering context.
    \item The script calls \texttt{toDataURL} method to extract the canvas image. 
    \item The script does not calls \texttt{save}, \texttt{restore}, and \texttt{addEventListener} methods on the canvas element.
    \end{enumerate}

\point{WebRTC Fingerprinting}
A script is identified as WebRTC fingerprinting script according to the following rules:
    \begin{enumerate}[leftmargin=0pt,itemindent=2em]
    \item The script calls \texttt{createDataChannel} or \texttt{createOffer} methods of the WebRTC peer connection.
    \item The script calls \texttt{onicecandidate} or \texttt{local Description} methods of the WebRTC peer connection.
    \end{enumerate}

\point{Canvas Font Fingerprinting}
A script is identified as canvas font fingerprinting script according to the following rules:
    \begin{enumerate}[leftmargin=0pt,itemindent=2em]
    \item The script sets the \texttt{font} property on a canvas element to more than 20 different fonts.
    \item The script calls the \texttt{measureText} method of the rendering context more than 20 times. 
    \end{enumerate}

\point{AudioContext Fingerprinting}
A script is identified as AudioContext fingerprinting script according to the following rules:
    \begin{enumerate}[leftmargin=0pt,itemindent=2em]
    \item The script calls any of the \texttt{createOscillator}, \texttt{createDynamicsCompressor}, \texttt{destination}, \texttt{start Rendering}, and \texttt{oncomplete} method of the audio context. 
    \end{enumerate}

\subsection{Examples of Dormant and Deviating Scripts}
\label{app:deviating-dormant}
Script \ref{lst:sample_dormant_script} shows an example dormant script and Script \ref{lst:sample_deviating_script} shows an example deviating script. 

\begin{figure}[htpb]
\begin{lstlisting}[style=htmlcssjs,caption=A truncated example of a dormant script from \protect\url{sdk1.resu.io/scripts/resclient.min.js} in which function prototypes are assigned to the \texttt{window} object and can be called at a later point in time., label={lst:sample_dormant_script}]
(function(g) {
......
n.prototype = {
  getCanvasPrint: function() {
  var b = document.createElement("canvas"),d;
  try {
      d = b.getContext("2d")
  } catch (e) {
      return ""
  }
  d.textBaseline = "top";
  d.font = "14px 'Arial'";
  ...
  d.fillText("http://valve.github.io", 4, 17);
  return b.toDataURL()
  }
};
"object" === typeof module &&
    "undefined" !== typeof exports && (module.exports = n);
g.ClientJS = n
})(window);
\end{lstlisting}
    \vspace{-25pt}
\end{figure}

\begin{figure}[htpb]
\begin{lstlisting}[style=htmlcssjs,caption=A truncated example of a deviating script from \protect\url{webresource.c-ctrip.com/code/ubt/_bfa.min.js?v=20195_22.js}. The heuristic is designed to ignore scripts that call \texttt{save} or \texttt{restore} on \texttt{CanvasRenderingContext2D} as a way to reduce false positives., label={lst:sample_deviating_script}]
...
canvas: function(t) {
var e = document.createElement("canvas");
if ("undefined" == typeof e.getContext) 
t.push("UNSUPPORTED_CANVAS");
else {
e.width = 780, e.height = 150;
var n = "UNICODE STRING",
i = e.getContext("2d");
i.save(), i.rect(0,0,10,10), i.rect(2,2,6,6),
t.push(!1 === i.isPointInPath(5, 5, "evenodd")
? "yes" : "no"), i.restore(), i.save();
var r = i.createLinearGradient(0, 0, 200, 0);
.....
i.shadowColor="rgb(85,85,85)",i.shadowBlur=3,
i.arc(500,15,10,0,2*Math.PI,!0),i.stroke(),
i.closePath(),i.restore(),t.push(e.toDataURL())
}
return t
}
...
\end{lstlisting}
    \vspace{-35pt}
\end{figure}

\subsection{Why Machine Learning?}
\label{sec:why-ml}
To conduct fingerprinting, websites often embed off-the-shelf third-party fingerprinting libraries.
Thus, one possible approach to detect fingerprinting scripts is to simply compute the textual similarity between the known fingerprinting libraries and the scripts embedded on a website. 
Scripts that have higher similarity with known fingerprinting libraries are more likely to be fingerprinting scripts. 
To test this hypothesis, we compare the similarity of fingerprinting and non-fingerprinting scripts detected by \framework against fingerprintjs2, a popular open-source fingerprinting library. 
Specifically, we tokenize scripts into keywords by first beautifying them and then splitting them on white spaces.
We then compute a tokenized script's Jaccard similarity, pairwise, with all versions of fingerprintjs2.
The highest similarity score among all versions is attributed to a script.

\begin{figure}[htbp]
    \centering
             \vspace{-10pt}
    \includegraphics[width=.5\textwidth]{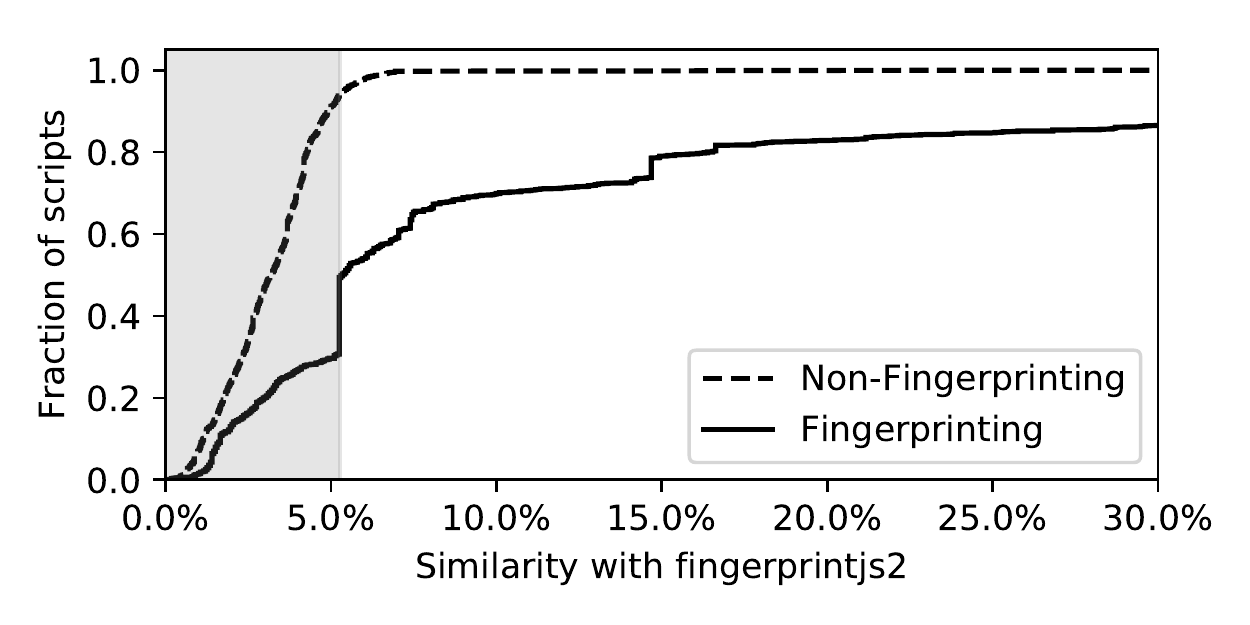}
         \vspace{-20pt}
    \caption{Jaccard similarity of fingerprinting and non-fingerprinting scripts with fingerprintjs2. The shaded portion of the figure represents the scripts classified as non-fingerprinting and the clear portion of the figure represents the scripts classified as fingerprinting based on the similarity threshold.}
    \label{figure:ml_simm}
     \vspace{-15pt}
\end{figure}

Our test set consists of the fingerprinting scripts detected by \framework and an equal number of randomly sampled non-fingerprinting scripts.
Figure~\ref{figure:ml_simm}, plots the similarity of \framework's detected fingerprinting and non-fingerprinting scripts with fingerprintjs2.
We find that the majority of the detected fingerprinting scripts (54.06\%) have less than 6\% similarity to fingerprintjs2 and only 13.49\% of the scripts have more than 30\% similarity.
Whereas most of the detected non-fingerprinting scripts (90.94\%) have less than 5\% similarity to fingerprintjs2 and only 9.05\% of the scripts have more than 5\% similarity.
We find that the true positive rate is at the highest (69.20\%) and false positive rate is at the lowest (5.97\%) with an accuracy of 81.69\%, when we set the similarity threshold to 5.28\%.
The shaded portion of the figure represents the scripts classified as non-fingerprinting and the clear portion of the figure represents the scripts classified as fingerprinting using this threshold. 
There is a significant overlap between the similarity of both fingerprinting and non-fingerprinting scripts and there is no optimal way to use similarity as a classification threshold.

Overall, our analysis shows that most websites do not integrate fingerprinting libraries as-is but instead make alterations.
Alterations often include embedding minified or obfuscated versions of the fingerprinting libraries, embedding only a subset of the fingerprinting functionality, or fingerprinting libraries inspired re-implementation.
Such alterations cause a lower similarity between fingerprinting scripts and popular fingerprinting libraries. 
We also find that several APIs are frequently used in both fingerprinting and non-fingerprinting scripts. 
Common examples include the use of utility APIs such as Math and window, and non-fingerprinting scripts using fingerprinting APIs for functional purposes e.g. canvas API being used for animations.
The presence of such APIs results in increase of similarity between non-fingerprinting scripts and fingerprinting libraries.
A simple similarity metric cannot generalize on alterations to fingerprinting libraries and functional uses of APIs, and thus fails to detect fingerprinting scripts. 
Whereas, our syntactic-semantic machine learning approach is able to generalize.
Our analysis justifies the efficacy of a learning based approach over simple similarity metric.

\subsection{JavaScript APIs Frequently Used in Fingerprinting Scripts}
\label{app:high-frequency-fp-apis}
Below we provide a list of JavaScript API keywords frequently used by fingerprinting scripts.
To this end, we measure the relative prevalence of API keywords in fingerprinting scripts by computing the ratio of their fraction of occurrence in fingerprinting scripts to their fraction of occurrence in non-fingerprinting scripts.
A higher value of the ratio for a keyword means that it is more prevalent in fingerprinting scripts than non-fingerprinting scripts. 
Note that $\infty$ means that the keyword is only present in fingerprinting scripts. 
Table~\ref{tab:frequency-of-keywords} includes keywords that have pervasiveness values greater than or equal to 16 and are present on 3 or more websites.

\xentrystretch{-0.13}
\bottomcaption{JavaScript API keywords frequently used in fingerprinting scripts, and their presence on 20K websites crawl. Scripts (count) represents the number of distinct fingerprinting scripts in which the keyword is used and Websites (count) represents the number of websites on which those scripts are embedded.\label{tab:frequency-of-keywords}}
\tablefirsthead{\toprule \bfseries Keywords & \bfseries Ratio & \bfseries Scripts & \bfseries Websites\\
\bfseries & \bfseries & (count) &  (count)\\ 
\midrule}
\tabletail{
\\ \midrule}
\tablelasttail{
\\ \bottomrule}
\centering
\fontsize{8}{7}\selectfont
\setlength\tabcolsep{4pt} 
\begin{xtabular}{l|rrr}
    onpointerleave & $\infty$ & 4 & 1366\\
    StereoPannerNode & $\infty$ & 1 & 1363\\
    FontFaceSetLoadEvent & $\infty$ & 1 & 1363\\
    PresentationConnection\\
    AvailableEvent & $\infty$ & 1 & 1363\\
    msGetRegionContent & $\infty$ & 1 & 1363\\
    peerIdentity & $\infty$ & 1 & 1363\\
    MSManipulationEvent & $\infty$ & 1 & 1363\\
    VideoStreamTrack & $\infty$ & 1 & 1363\\
    mozSetImageElement & $\infty$ & 1 & 1363\\
    requestWakeLock & $\infty$ & 1 & 174\\
    audioWorklet & $\infty$ & 3 & 8\\
    onwebkitanimationiteration & $\infty$ & 3 & 3\\
    onpointerenter & $\infty$ & 3 & 3\\
    onwebkitanimationstart & $\infty$ & 3 & 3\\
    onlostpointercapture & $\infty$ & 3 & 3\\
    ongotpointercapture & 362.52 & 3 & 3\\
    onpointerout & 362.52 & 3 & 3\\
    onafterscriptexecute & 217.51 & 18 & 1380\\
    channelCountMode & 199.03 & 28 & 39\\
    onpointerover & 181.26 & 3 & 3\\
    onbeforescriptexecute & 181.26 & 18 & 1380\\
    onicegatheringstatechange & 179.78 & 61 & 61\\
    MediaDevices & 161.12 & 4 & 1366\\
    numberOfInputs & 157.09 & 26 & 36\\
    channelInterpretation & 147.69 & 11 & 22\\
    speedOfSound & 140.98 & 7 & 11\\
    dopplerFactor & 140.98 & 7 & 11\\
    midi & 138.72 & 225 & 251\\
    ondeviceproximity & 131.35 & 25 & 282\\
    HTMLMenuItemElement & 121.40 & 218 & 244\\
    updateCommands & 120.84 & 1 & 1363\\
    exportKey & 105.97 & 57 & 57\\
    onauxclick & 90.63 & 3 & 3\\
    microphone & 90.43 & 223 & 250\\
    iceGatheringState & 90.30 & 68 & 1481\\
    ondevicelight & 88.31 & 19 & 36\\
    renderedBuffer & 87.17 & 189 & 439\\
    WebGLContextEvent & 82.52 & 28 & 44\\
    ondeviceorientationabsolute & 80.56 & 4 & 1366\\
    startRendering & 79.33 & 193 & 458\\
    createOscillator & 78.77 & 191 & 445\\
    knee & 76.65 & 170 & 419\\
    OfflineAudioContext & 74.68 & 199 & 721\\
    timeLog & 72.50 & 12 & 12\\
    getFloatFrequencyData & 72.50 & 6 & 10\\
    WEBGL\_compressed\_texture\_atc & 72.50 & 3 & 4\\
    illuminance & 72.50 & 3 & 3\\
    reduction & 69.64 & 170 & 419\\
    modulusLength & 69.39 & 58 & 58\\
    WebGL2RenderingContext & 68.71 & 29 & 30\\
    enumerateDevices & 64.12 & 208 & 666\\
    AmbientLightSensor & 63.60 & 10 & 267\\
    attack & 61.31 & 173 & 434\\
    AudioWorklet & 60.42 & 22 & 32\\
    Worklet & 60.42 & 22 & 32\\
    AudioWorkletNode & 60.42 & 22 & 32\\
    lastStyleSheetSet & 60.42 & 1 & 1363\\
    DeviceProximityEvent & 60.42 & 1 & 1363\\
    DeviceLightEvent & 60.42 & 1 & 1363\\
    enableStyleSheetsForSet & 60.42 & 1 & 1363\\
    UserProximityEvent & 60.42 & 1 & 1363\\
    mediaDevices & 60.03 & 230 & 850\\
    vendorSub & 56.17 & 251 & 1728\\
    setValueAtTime & 55.29 & 167 & 417\\
    getChannelData & 55.18 & 195 & 460\\
    MAX\_DRAW\_BUFFERS\_WEBGL & 54.93 & 10 & 12\\
    reliable & 52.36 & 39 & 103\\
    WEBGL\_draw\_buffers & 52.09 & 25 & 27\\
    EXT\_sRGB & 51.79 & 3 & 4\\
    setSinkId & 50.35 & 5 & 1367\\
    namedCurve & 50.29 & 67 & 74\\
    WEBGL\_debug\_shaders & 45.31 & 3 & 4\\
    productSub & 42.79 & 734 & 2819\\
    hardwareConcurrency & 41.92 & 716 & 3661\\
    publicExponent & 41.52 & 67 & 74\\
    requestMIDIAccess & 40.28 & 1 & 1363\\
    mozIsLocallyAvailable & 40.28 & 1 & 174\\
    ondevicemotion & 40.28 & 4 & 4\\
    XPathResult & 39.73 & 218 & 417\\
    mozBattery & 39.04 & 42 & 322\\
    IndexedDB & 38.73 & 25 & 25\\
    generateKey & 37.46 & 62 & 62\\
    buildID & 36.52 & 272 & 414\\
    getSupportedExtensions & 36.46 & 534 & 1007\\
    MAX\_TEXTURE\_MAX\_\\
    ANISOTROPY\_EXT & 35.85 & 521 & 980\\
    oscpu & 35.33 & 681 & 1196\\
    oninvalid & 34.75 & 65 & 1428\\
    vpn & 34.53 & 24 & 24\\
    createDynamicsCompressor & 33.54 & 189 & 442\\
    privateKey & 33.46 & 67 & 74\\
    EXT\_texture\_filter\_anisotropic & 32.91 & 479 & 949\\
    isPointInPath & 32.17 & 481 & 949\\
    getContextAttributes & 31.76 & 460 & 920\\
    BatteryManager & 31.23 & 23 & 50\\
    getShaderPrecisionFormat & 30.81 & 450 & 915\\
    depthFunc & 30.81 & 452 & 921\\
    uniform2f & 30.71 & 460 & 930\\
    rangeMax & 30.36 & 449 & 902\\
    rangeMin & 30.24 & 446 & 897\\
    EXT\_disjoint\_timer\_query & 30.21 & 3 & 4\\
    scrollByPages & 30.21 & 1 & 1363\\
    CanvasCaptureMediaStreamTrack & 30.21 & 1 & 18\\
    onlanguagechange & 30.21 & 4 & 4\\
    clearColor & 29.16 & 457 & 916\\
    createWriter & 28.93 & 17 & 17\\
    getUniformLocation & 28.61 & 466 & 948\\
    getAttribLocation & 28.58 & 464 & 945\\
    drawArrays & 28.53 & 466 & 948\\
    useProgram & 28.37 & 467 & 949\\
    enableVertexAttribArray & 28.37 & 466 & 948\\
    createShader & 28.31 & 467 & 949\\
    compileShader & 28.30 & 467 & 936\\
    shaderSource & 28.27 & 466 & 936\\
    attachShader & 28.25 & 464 & 934\\
    bufferData & 28.24 & 466 & 938\\
    linkProgram & 28.23 & 464 & 933\\
    vertexAttribPointer & 28.22 & 464 & 933\\
    bindBuffer & 28.14 & 463 & 932\\
    createProgram & 27.95 & 464 & 934\\
    OES\_standard\_derivatives & 27.46 & 20 & 1384\\
    appCodeName & 27.03 & 325 & 1890\\
    getAttributeNodeNS & 26.49 & 16 & 21\\
    ARRAY\_BUFFER & 25.36 & 471 & 941\\
    suffixes & 25.14 & 775 & 1441\\
    TouchEvent & 25.01 & 481 & 1130\\
    MIDIPort & 24.17 & 2 & 19\\
    onaudioprocess & 23.64 & 9 & 17\\
    showModalDialog & 23.56 & 39 & 1419\\
    globalStorage & 23.48 & 245 & 1681\\
    camera & 22.76 & 229 & 255\\
    onanimationiteration & 22.66 & 3 & 3\\
    textBaseline & 21.76 & 888 & 3234\\
    MediaStreamTrackEvent & 21.32 & 3 & 1365\\
    deviceproximity & 21.13 & 25 & 26\\
    taintEnabled & 20.89 & 14 & 24\\
    alphabetic & 20.65 & 671 & 2986\\
    userproximity & 20.28 & 24 & 25\\
    globalCompositeOperation & 20.15 & 507 & 975\\
    outputBuffer & 20.14 & 12 & 34\\
    WebGLUniformLocation & 20.14 & 1 & 1363\\
    WebGLShaderPrecisionFormat & 20.14 & 1 & 1363\\
    createScriptProcessor & 20.14 & 11 & 20\\
    createBuffer & 19.98 & 472 & 954\\
    UIEvent & 19.93 & 47 & 63\\
    toSource & 19.54 & 416 & 2224\\
    createAnalyser & 19.33 & 12 & 17\\
    fillRect & 19.22 & 898 & 3432\\
    evenodd & 18.49 & 504 & 960\\
    fillText & 18.09 & 957 & 3502\\
    candidate & 18.03 & 178 & 1847\\
    WEBGL\_debug\_renderer\_info & 17.83 & 406 & 2214\\
    toDataURL & 17.64 & 951 & 3507\\
    dischargingTime & 17.53 & 38 & 54\\
    bluetooth & 17.28 & 225 & 424\\
    FLOAT & 16.89 & 467 & 939\\
    battery & 16.82 & 152 & 1853\\
    devicelight & 16.51 & 25 & 26\\
    onanimationstart & 16.48 & 3 & 3\\
    getExtension & 16.43 & 575 & 1115\\
    onemptied & 16.11 & 4 & 4\\
\end{xtabular}

\end{document}